\documentclass[twocolumn,reprint]{aastex631}

\usepackage{amsmath,amssymb,graphicx,esint}
\usepackage{array}
\newcolumntype{P}[1]{>{\centering\arraybackslash}p{#1}}

\begin{document}

\title{Mass Cycle and Dynamics of a Virtual Quiescent Prominence}

\author[0009-0002-3036-5951]{Dion Donn\'e}
\affiliation{Centre for mathematical Plasma-Astrophysics, KU Leuven, Leuven, Belgium}
\email{dion.donne@kuleuven.be}

\author[0000-0003-3544-2733]{Rony Keppens}
\affiliation{Centre for mathematical Plasma-Astrophysics, KU Leuven, Leuven, Belgium}



\begin{abstract}
The mass cycle of solar prominences or filaments is still not completely understood. Researchers agree that these dense structures form by coronal in-situ condensations and plasma siphoning from the underlying chromosphere. In the evaporation-condensation model siphoning arises due to evaporation of chromospheric plasma from localised footpoint heating but this is challenging to justify observationally. Here, we simulate the reconnection-condensation model at extreme-resolutions down to 20.8 km within a three-dimensional magnetohydrodynamic coronal volume. We form a draining, quiescent prominence and associated coronal rain simultaneously. We show that thermal instability – acting as a trigger for local condensation formation – by itself drives siphoning flows from the low-corona without the need of any localised heating. In addition, for the first time we demonstrate through a statistical analysis along more than 1000 magnetic field lines that cold condensations give rise to siphoning flows within magnetic threads.  This siphoning arises from the strong pressure gradient along field lines induced by thermal instability.  No correlation is found between siphoning flows and the prominence mass, making thermal instability the main in-situ mass collection mechanism.  Our simulated prominence drains by gliding along strongly sheared, asymmetric, dipped magnetic arcades, and develops natural vertical fine-structure in an otherwise horizontal magnetic field due to the magnetic Rayleigh-Taylor instability. By synthesising our data, our model shows remarkable agreement with observations of quiescent prominences such as its dark coronal cavity in extreme-ultraviolet emission channels, fine-scale vertical structure and reconnection outflows which, for the first time, have been self-consistently obtained as the prominence evolves.
\end{abstract}



\section{Introduction} \label{sec:intro}

Solar prominences, or solar filaments, are suspended plasma condensations in the solar corona that are roughly one hundred times colder and denser than their ambient environment. While these structures have been discovered for more than a century \citep{secchi1875}, many questions still remain. Their formation process is still open to debate but it is generally believed that the local plasma condenses due to thermal instability or catastrophic cooling \citep{parker1953,field1965}. While mere in-situ condensations due to catastrophic cooling can realize the lower limit of observed mass values in solar prominences, it is challenging to explain the mass values that heavy solar prominences inherit since the solar corona does not possess enough mass for condensations to reach average values of $10^{14}-10^{15} \,  \mathrm{g}$ \citep{zirker1994}. Therefore, siphoning flows from the chromosphere are theorised to overcome this mass discrepancy. These siphoning flows form the basis of all numerical models that adopt the evaporation-condensation mechanism which gradually evaporates upper chromospheric matter into the corona \citep{poland1986}. In practice, a localised heating term in the governing magnetohydrodynamic (MHD) energy equation is added to evaporate the plasma from denser low-lying regions and this elevated matter subsequently condenses at the top of the (dipped) magnetic arcades in 1D \citep{xia2011, zhou2014} and 2D \citep{xia2012, keppens2014, jervcic2022, zhou2023}. The evaporation-condensation model has also been implemented for 3D flux rope models \citep{xia2016}. In such foot-point heated scenarios, the heating length scale of the localised heating source should not be too large compared to the length of the magnetic field line in order not to heat up and evaporate the solar prominence \citep{mackay2021}. Only recently have such chromospheric siphoning flows, induced by magnetic reconnection at the footpoints, been detected in the context of solar filament formation  \citep{yang2021}. In contrast, the reconnection-condensation model does not incorporate chromospheric heating \citep{kaneko2017}, and was shown to lead to a 3D prominence structure in a coronal-only setting. It emphasizes how lower-lying but still coronal matter in a stratified medium is scooped up during reconnection-induced flux rope formation which then undergoes thermal instability and condenses into a solar prominence. Here, we will revisit this latter reconnection-condensation model at extreme resolution, and for the first time fully analyse its mass cycle and finer-scale structuring.

Coronal rain has overall the same plasma properties as a solar prominence, except that these smaller and more dynamic condensations are not suspended by the magnetic field but rather fall downwards towards the solar surface.  They have been observed in the solar corona \citep{antolin2010,sahin2023} and studied numerically as well in 1D \citep{antolin2010, luna2012}, 2D \citep{fang2013, fang2015, li2022} and 3D \citep{moschou2015, xia2017, kohutova2020}, but these simulations have only been carried out for magnetic arcades. Also for these condensations, the ultimate local trigger is thermal instability to form dense and cold blobs.  They attain a variety of speeds between $0 - 150  \,  \mathrm{km} \, \mathrm{s}^{-1}$,  depending on their initial height \citep{antolin2012}.  When they are numerically simulated, localised heating at the footpoints is a recurring model ingredient \citep{fang2013, li2022}. The parameter space of localised heating has been researched through many 1D studies \citep{muller2003, muller2005, antolin2010, pelouze2022}. Recently, few studies have departed from the traditional, anomalous footpoint heating method and instead developed a self-consistent model for chromospheric evaporation by enabling magnetoconvection using the \texttt{Bifrost} code \citep{bifrost} in 2.5D \citep{antolin2022} and 3D \citep{kohutova2020} to form coronal rain. 

Not only the formation process of solar prominences is not completely understood but self-consistently modeling their microscopic dynamics poses challenges as well. The morphologies of solar prominences exhibit vertical structure and movement in mostly horizontal magnetic fields as can clearly be seen in observations \citep{berger2008}.  The literature advocates that the magnetic Rayleigh-Taylor instability (mRTI) is the prime cause for naturally developing such vertical structures \citep{berger2010,  terradas2015,  xia2016, kaneko2018}. Recently, \citet{jenkins2022} simulated the in-situ formation of a 3D MHD solar prominence at extreme resolution (20 km) and concluded that the interchange mode of the mRTI, i.e.  perturbations that are oriented perpendicular to the local magnetic field,  dominates the prominence evolution. Their work focused on a pure flux-rope magnetic structure, within which a small prominence condensed in-situ and evolved through mRTI. In this work, we will use a reconnection-condensation route to form a mixed arcade plus flux rope magnetic structure, which is much more realistic for actual prominence settings.

In this work, a quiescent solar prominence and coronal rain form simultaneously in the reconnection-condensation model and without localised footpoint heating. We fully elucidate the mass cycle and internal dynamics of the quiescent solar prominence and relate it to observations through synthesised images.  Therefore, the paper is structured as follows.  Section 2 focuses on the setup of our simulation where we highlight the sink- and source terms in the MHD equation and how we are able to achieve an extreme resolution down to 20.8 km.  Section 3 is split in four subsections where we provide: an in-depth overview on the formation of the solar prominence and coronal rain,  a quantitative study of the mass cycle,  the dynamics of mRTI and finally our synthesised data (with added movies).  We end our findings with a summary in Section 4. 

\section{Simulation Setup}
We simulate a large domain of $ -20 \, \mathrm{Mm} \leqslant x \leqslant 20 \, \mathrm{Mm}$,  $ 0 \, \mathrm{Mm} \leqslant y \leqslant 160 \, \mathrm{Mm}$ and $ -100 \, \mathrm{Mm} \leqslant z \leqslant 100 \, \mathrm{Mm}$ where the vertical direction is $y$. This large size is adopted to prevent the solar prominence from interacting with the boundaries, and prepare for a future study of prominence eruption onsets.  

Our initial condition follows the setup of \citet{kaneko2017} closely.  It starts with an isothermal coronal atmosphere at $T_0 = 1\, \mathrm{MK}$.  The density profile in typical solar external gravity $g = -274 \,  \mathrm{m} \, \mathrm{s}^{-1}$ then becomes 
\begin{align}
\rho &= \rho_0 \exp\bigg(-\dfrac{y}{H_0} \bigg),
\end{align}
with $H_0 = k_\mathrm{B} T_0 / (m_\mathrm{plasma} g )$ the scale height, where $k_\mathrm{B}$ is Boltzmann's constant.   Because we model a fully ionised mixture with a helium-to-hydrogen number ratio of $n_\mathrm{He}/n_\mathrm{H} = 0.1$,  we have the $m_\mathrm{plasma} = 0.6 m_\mathrm{p}$ such that $H_0 = 50\, \mathrm{Mm}$.  At $y=0 \,  \mathrm{Mm} $ we take a number density of $n_0=10^9 \, \mathrm{cm}^{-3}$ to modulate the low coronal environment which results in a density value of $\rho_0 = 2.3 \times 10^{-15} \,\mathrm{g} \, \mathrm{cm}^{-3}$.   The pressure is then retrieved from the ideal gas law.

The initial magnetic field consists of force-free magnetic arcades, such that total force-balance is achieved with a hydrostatic atmosphere.  The magnetic field component variation of the arcades are
\begin{align}
B_x & =-\left(\frac{2 L_\mathrm{a}}{\pi H_0}\right) B_0 \cos \left(\frac{\pi x}{2 L_\mathrm{a}}\right) \exp \left[-\frac{y}{H_0}\right] \,,\\
B_y & =B_0 \sin \left(\frac{\pi x}{2 L_\mathrm{a}}\right) \exp \left[-\frac{y}{H_0}\right] \,,\\
B_z & =-\sqrt{1-\left(\frac{2 L_\mathrm{a}}{\pi H_0}\right)^2} B_0 \cos \left(\frac{\pi x}{2 L_\mathrm{a}}\right) \exp \left[-\frac{y}{H_0}\right]\,,
\end{align}
with $B_0 = 8\,\mathrm{G}$ and $L_\mathrm{a} = 20\,\mathrm{Mm}$.  Note that the magnetic field strength decays with the same scale height $H_0$ as the density and pressure,  i.e.  $\vert \mathbf{B} \vert = B_0 \exp(-y/H_0)$.

The full 3D MHD equations formulated for the conserved variables with non-adiabatic and resistive source terms are the following \citep{goedbloed2019}:
\begin{align}
& \frac{\partial \rho}{\partial t}+\nabla \cdot(\rho \mathbf{v})=0,\label{eq:continuity}\\
& \frac{\partial}{\partial t}(\rho \mathbf{v})+\nabla \cdot\bigg[\rho \mathbf{v v}+\bigg(p+\frac{1}{2} B^2\bigg) \mathbf{I}-\mathbf{B B}\bigg]=\rho \mathbf{g}, \\
& \frac{\partial \mathbf{B}}{\partial t}+\nabla \cdot(\mathbf{v B}-\mathbf{B v})=-\nabla \times(\eta \mathbf{j}), \label{eq:induction} \\
& \frac{\partial}{\partial t}\bigg(\frac{1}{2} \rho v^2+\rho e+\frac{1}{2} B^2\bigg) \nonumber \\
&+\nabla \cdot\bigg[\bigg(\frac{1}{2} \rho v^2+\rho e+p\bigg) \mathbf{v}+ (-\mathbf{v} \times \mathbf{B}+\eta \mathbf{j}) \times \mathbf{B}\bigg]=  \nonumber \\
&\qquad\rho \mathbf{v} \cdot \mathbf{g} -n^2 \Lambda(T)+H  +\nabla \cdot\bigg(\kappa_{\|} T^{5 / 2} \mathbf{b b} \cdot \nabla T\bigg), \\
&\nabla \cdot \mathbf{B}=0.
\end{align}
In the equations,  we solve for the density $\rho$,  velocity $\mathbf{v}$,  magnetic field $\mathbf{B}$ and the internal energy $e = p/(\gamma\rho - \rho)$ where we already adopted units where magnetic permeability $\mu_0=1$.

Our right-hand-side source terms are the following. We use anomalous resistivity $\eta$ which activates above a critical local current magnitude value as prescribed by \citet{kaneko2017},  i.e.
\begin{align}
\eta&=0,  &\left(J<J_\mathrm{c}, t \geqslant t_2\right) \\
\eta&=\eta_0\left(J / J_\mathrm{c}-1\right)^2,  &\left(J \geqslant J_\mathrm{c}\right)
\end{align}
with $\eta_0 = 3.6 \cdot 10^{13} \, \mathrm{cm}^2 \, \mathrm{s}^{-1}$, $J_\mathrm{c} = 25 \, \mathrm{erg}^{1/2} \,  \mathrm{cm}^{-3/2}\, \mathrm{s}^{-1}$ and $t_2 = 4\,500 \, \mathrm{s}$. Note that after $t = t_2$, anomolous resistivity is turned off, though numerical resistivity will still inevitably be present. Maximum resistivity is set at an upper limit of $\eta_\mathrm{max} = 1.8 \cdot 10^{14} \, \mathrm{cm}^2 \, \mathrm{s}^{-1}$. The volumetric radiative cooling rate $\Lambda(T)$ is interpolated from the values of the \texttt{Colgan\_DM} radiative cooling curve for which we have assumed an optically thin medium \citep{hermans2021}.  To prevent the coronal atmosphere from only being cooled,  a static heating term is included to maintain an energy-balanced initial condition,  i.e.  $H = n_0^2 \Lambda(T_0) \exp(-2y / H_0)$. It is to be noted that this background heating term is not sufficient for triggering condensations in simulations where the chromospheric regions are included as well. Indeed, they all require a substantially strong, additional localized heating which we do not use in this work. Furthermore, \citet{brughmans2022} have conducted a parameter study on how different prescriptions of this background heating rate $H$ affect the final properties of the condensations with 2.5D simulations. They showed that a reduction of the background heating within the flux rope naturally leads to more volume-filling condensation formations throughout the flux rope. But a static exponential is still sufficient in promoting condensations.

Lastly,  anisotropic thermal conduction is used where the parallel magnetic field conductivity is modelled according to the Spitzer conductivity with $\kappa_\parallel = 8\times 10^{-7} \, \mathrm{erg} \,  \mathrm{cm}^{-1} \, \mathrm{s}^{-1} \, \mathrm{K}^{-1}$ \citep{spitzer2006}.

In addition to an initial condition,  the MHD equations need to be supplied with boundary conditions as well.  The density and pressure of the ghostcells in the bottom boundary $y<0 \,  \mathrm{Mm} $ follow the same isothermal,  hydrostatic profile as during the initial condition, i.e.  $\rho = \rho_0 \exp\big(-y/H_0 \big)$ and $p = k_\mathrm{B} \rho T_0 / m_\mathrm{plasma}$. Only in the central region of our domain and for a finite time, the footpoints of the magnetic arcades converge towards the central $x=0$ polarity inversion line (PIL) with velocity $v_x(x,z,t) = -v_0(t) \sin \big(\pi x /L_\mathrm{a}\big) \exp\big( -z^2 / L_\mathrm{a}^2 \big)$ and $v_y = v_z = 0$ where
\begin{equation}
v_0(t) =\left\{\begin{array}{lll}
v_{00} & \text { if } t < t_1 \\
v_{00} \frac{t_2-t}{t_2-t_1} & \text { if } t_1 \leq t < t_2 & \\
0 & \text { else } \\
\end{array}\right. 
 \end{equation}
 with $v_{00} = 6 \, \mathrm{km} \, \mathrm{s}^{-1}$,  $t_1 = 4\,320 \, \mathrm{s}$ and $t_2 = 4\,500\, \mathrm{s}$.  Note that although $v_y$ is set to zero in the centers of the ghostcells $y \leqslant 0 \,  \mathrm{Mm}$,   a non-zero flux is still able to enter and leave the boundary $y = 0 \, \mathrm{Mm}$. This is because our finite-volume, shock-capturing treatment computes fluxes at the boundary through a two-sided evaluation employing limited, non-linear reconstructions from cell-center to cell-edge values. Hence, the bottom boundary is at least partially open, and this aspect will be important for our analysis of siphoning flows later on.
The magnetic field at the bottom is calculated according to a zero-gradient second order extrapolation from the inner domain
 \begin{equation}
 \mathbf{B}(y_i) = \big( 4\mathbf{B}(y_{i+1}) - \mathbf{B}(y_{i+2}) \big) /3,
 \end{equation}
similarly as in the numerical setup of \citet{jenkins2021}. For all the five other boundaries,  the density of the ghostcells is copied from the bordering, inner cell.  The pressure is then calculated from the same ideal gas law equation by enforcing the initial coronal temperature $T_0=1\,\mathrm{MK}$.  The magnetic field is calculated from the zero-gradient second order extrapolation. The velocity field is nullified except for the top boundary. There,  the velocity field is calculated from the zero-gradient second order extrapolation and has the added constraint that the vertical velocity at the top $v_y$ should be strictly positive to prevent artificial inflow from occurring. In practice, this allows outflows and partially open conditions at top and bottom coronal boundaries, while all side boundaries keep their near-initial coronal conditions.

We solve the MHD equations with our open-source code \texttt{MPI-AMRVAC} \citep{amrvac3.0}.  The base resolution is compounded of $60 \times 240 \times 300$ cells and we use at least four additional refinement levels,  reaching an effective resolution of $\approx 41.7 \, \mathrm{km}$. A simulation where an additional level is activated to reach about 20.8 km cell sizes is discussed as well.  The  MHD equations are solved using the \texttt{HLL} method,  a \texttt{fourstep} timestepper and the \texttt{vanleer} limiter. To maintain a divergenceless magnetic field,  we make use of the \texttt{linde} method that adds a parabolic source term in the MHD equations to diffuse monopole errors. We refer the reader to the recent code paper \citep{amrvac3.0} for full references to the original papers that introduced these various algorithmic strategies.

Note that our effective resolution is maximally $1920\times 7680 \times 9600$, which is extreme, and this is only doable thanks to our grid-adaptive strategy. The original reconnection-condensation simulation \citep{kaneko2017} adopted a uniform grid of order $200\times 330 \times 540$, with 120 km cell resolution. Our entire domain is significantly larger than their $24 \, \mathrm{Mm}\times 40 \, \mathrm{Mm} \times 65 \, \mathrm{Mm}$ box, which will allow us to do future follow-up studies of prominence eruptions. Due to the rapid increase in actual finest-level cells in 3D, additional constraints on the adaptive mesh refinement (AMR) algorithm are imposed to maintain computational affordability.  Cells that lie inside the most central region $\vert x \vert \leqslant 7.5 \, \mathrm{Mm}$,  $y \leqslant 25 \, \mathrm{Mm}$ and $\vert z \vert \leqslant 40 \, \mathrm{Mm}$ are resolved with at least two refinement levels, such that this entire region is covered by cells of 166 km or finer. To follow all fine-structured condensations in their runaway cooling phase, those cells within the most central region that satisfy $T\leqslant 0.1 \, \mathrm{MK}$ in addition are captured at the maximum resolution. Cells that lie outside this central region, even if those cells happen to contain condensations, are captured at the coarsest resolution, i.e.  $\approx 667 \, \mathrm{km}$. This setting ensures that a high computational speed is maintained during the formation of the flux rope and that high-resolution refinement is activated well in advance before the first signs of plasma condensations occur.   

By exploiting the above refinement criteria and combining it with our numerical solution methods, our simulation required about two weeks of calculation time on more than $3\,000$ CPUs (on average) in order to evolve the MHD equations up to $9\,000 \, \mathrm{s}$. We developed a workflow to visualise the entire 3D domain in an interactive manner, explore isosurfaces of various parameters, combined with information from 1D lines, 2D slices and 3D volumes into a single frame. We automated acquisition of 1D data (for example extracting values of MHD variables along any number of vectors with any orientation and location at any time) and isolating specific fine-structure through zooming in and excluding values outside a customisable parameter range.

 \begin{figure*}[!t]
\begin{center}
\includegraphics[width=\linewidth]{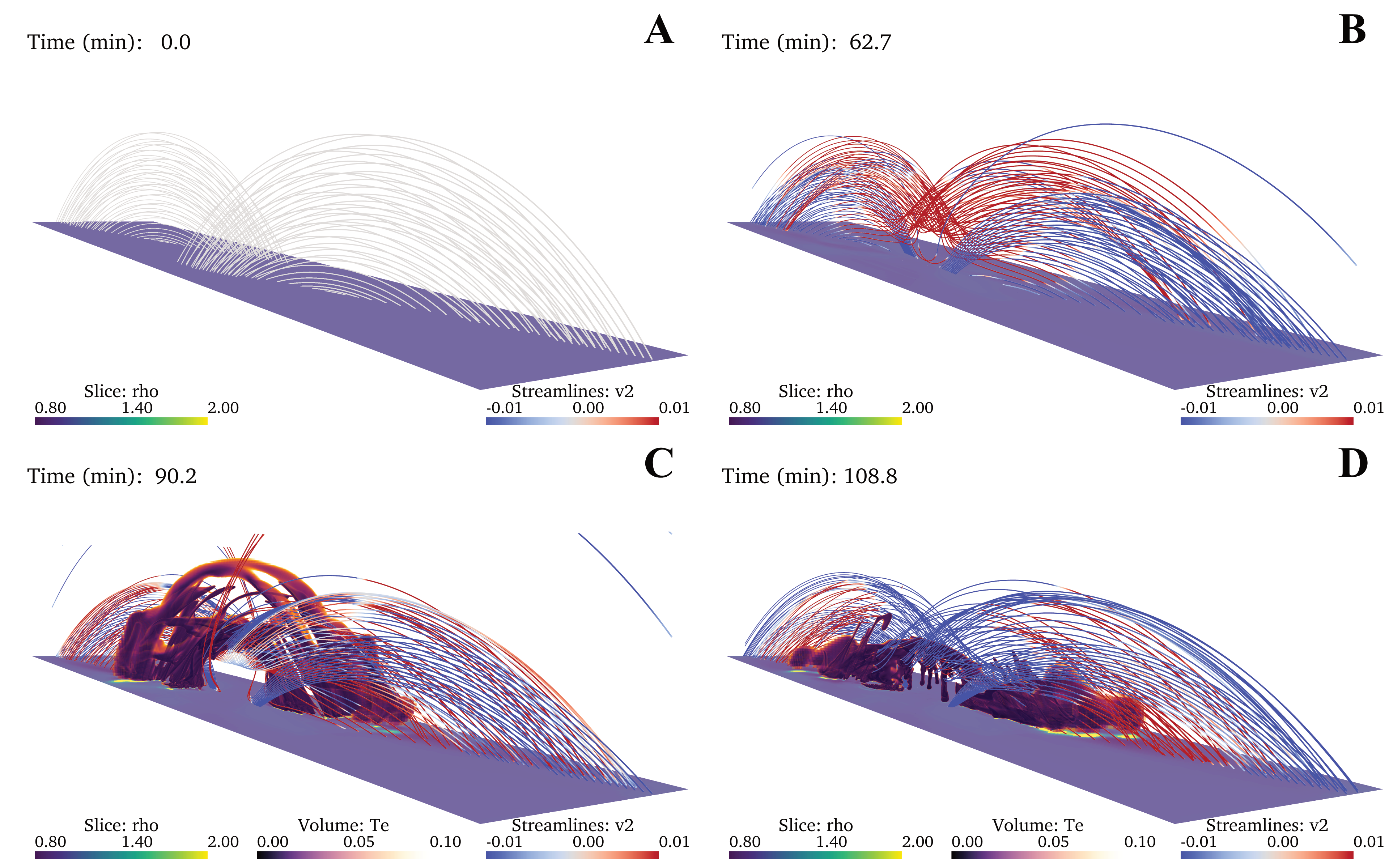}
 \end{center}
 \caption{The formation of a solar prominence and associated coronal rain. The bottom plane displays the density $\rho$ [in units of $2.3 \cdot 10^{-15} \, \mathrm{g} \, \mathrm{cm}^{-3}$], while the deforming magnetic field lines are colored by the vertical velocity of the plasma $v_y $ [in units of $116 \, \mathrm{km} \, \mathrm{s}^{-1}$]. From Panel \textbf{C} onwards, a volume rendering exhibits those cells in the range $T \leqslant 0.1 \, \mathrm{MK}$.  The colours of $\rho$ and $v_y$ are saturated in order to manifest contrasts.  \textbf{A}) The initial magnetic arcades at $t=0 \, \mathrm{min}$.  \textbf{B})  As the footpoints of the opposing magnetic arcades approach each other,  they reconnect due to resistivity $\eta$,  altering the magnetic topology into a flux rope.  \textbf{C}) First condensations of prominence and coronal rain occur soon after the flux rope has formed.  \textbf{D}) As the coronal rain falls towards the bottom and leaves the system, only the prominence matter is left, showing Rayleigh-Taylor fingering signatures in the central part of the filament. The entire simulation domain is visualised except for the height which is constrained to $y \leqslant 35 \, \mathrm{Mm}$. The animation of this figure lasts for 21 seconds which manifests the evolution of the prominence formation and the onset of the mRTI.  (An animation of this figure is available in the journal's webpage of this paper.)} \label{fig:promform}
 \end{figure*}

\subsection*{Formation of the Solar Prominence and Coronal Rain}
Fig.~\ref{fig:promform} exhibits the evolution of our simulation through four snapshots. The initial magnetic arcades in the isothermal, hydrostatic atmosphere are shown in Fig.~\ref{fig:promform}A. As the footpoints of the magnetic arcades with opposite magnetic polarities approach one another in the central region of our domain,  the local current density $\mathbf{J}$ increases significantly, resulting in the local resistivity coefficient attaining its maximum value $\eta_\mathrm{max}$. Hence, a local current sheet is formed within which magnetic field lines reconnect. Since the current sheet is fed with magnetic field lines through its sides, whose vertical orientation is opposing, the reconnected magnetic field lines are driven upwards, allowing a central flux rope structure to form and the magnetic flux rope to rise,  Fig.~\ref{fig:promform}B. The denser plasma from the lower coronal region in the stratified medium is scooped up during the flux rope formation,  undergoing thermal instability and condensing into a solar prominence, as seen in Fig.~\ref{fig:promform}C. In an animated view of this volume-rendered condensed material, it can be seen that in addition to a solar prominence, coronal rain occurs as well. As the system evolves further, the coronal rain falls towards the bottom since it is not trapped within the flux rope and proceeds to sink through our partially open bottom boundary. From Fig.~\ref{fig:promform}D and in closer-up views which we will analyse further on, it can be seen that the prominence mass exhibits signatures of a magnetic Rayleigh-Taylor instability. This is consistent with other 3D MHD simulations \citep{terradas2015, xia2016,  jenkins2022}, but note that here we find it to occur ab-initio, during and after formation of the prominence body. The fact that magnetic Rayleigh-Taylor instabilities are observed across a variety of magnetic topologies hosting prominences shows that this is an intrinsic phenomenon in solar prominences. Note that the central regions where the mRTI fingers develop clearly have locally upward concave magnetic field lines, with an overall horizontal orientation as well. In the animation of Fig.~\ref{fig:promform} it can be seen that plasma is escaping the flux rope by gliding along the magnetic field lines towards the bottom, parallel to the spine of the solar prominence. From Figs.~\ref{fig:promform}C and \ref{fig:promform}D we note that the end of the magnetic field lines have a positive upward plasma velocity (which can be seen by the red colouring), exhibiting an inflow of matter. We will return to this flow pattern and its significance for the prominence mass cycle in following sections.

\begin{figure*}[!t]
\includegraphics[width=\textwidth]{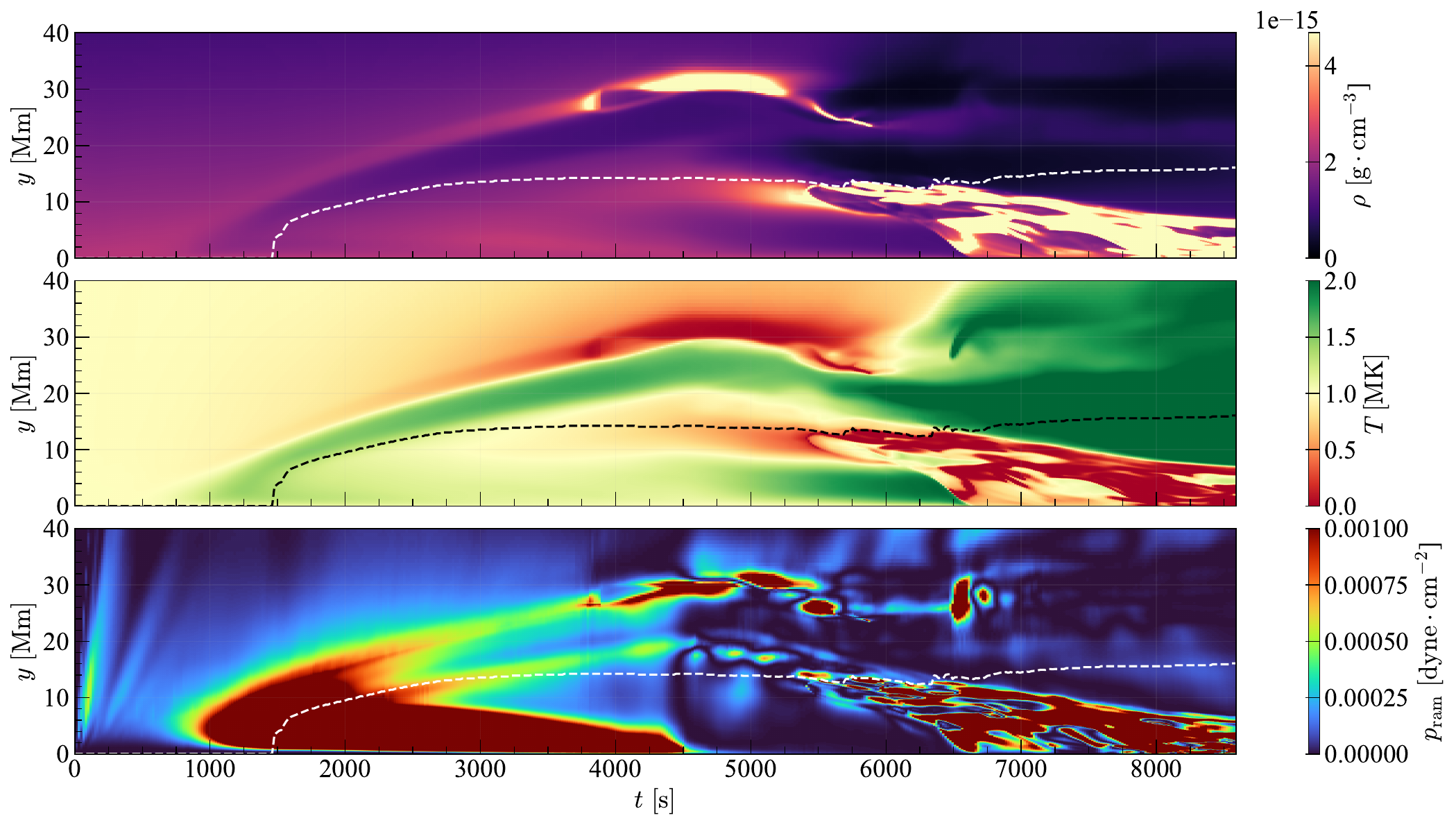}
\caption{Evolution of a 1D vertical ray intersecting with the solar prominence material and with coronal rain. The white dashed line traces the location of the flux rope's centre.  The density $\rho$, temperature $T$ and vertical ram pressure $p_\mathrm{ram}$ are expressed in CGS-units. Colorbars are saturated in order to highlight contrasts.}\label{fig:promform_1D_evolve}
\end{figure*}

The dynamics of the flux rope and the formation of the solar prominence and coronal rain can be better understood when a 1D vertical along the $y$-axis is taken through the center of the domain,  intersecting with these two types of condensations. The evolution of the density $\rho$, temperature $T$ and ram pressure $p_\mathrm{ram} = \rho v_y^2$ as a function of time along this vertical is shown in Fig.~\ref{fig:promform_1D_evolve}. The figure also shows the location of the flux rope's centre, represented by the white dashed line. The flux rope starts to form at $t \approx 750 \, \mathrm{s}$ due to magnetic reconnection. From the bottom panel, the energy release from magnetic reconnection exerts a high upwards ram pressure on the flux rope,  pushing it upwards until the flux rope's centre reaches an altitude of $\sim 16 \, \mathrm{Mm}$.  At $t = 4\,500 \,  \mathrm{s}$, footpoint motion in the bottom boundary is turned off and thus magnetic reconnection under the flux rope ceases to exist as well, such that from that time we witness a vanishing vertical ram pressure.  While the flux rope rises in altitude between $t=750 \, \mathrm{s}$ and $t=3\,750 \, \mathrm{s}$, its top bounding surface rises as well and the associated gas and magnetic pressure pushes on the surrounding medium, increasing the local density and ram pressure as can be seen at $y\approx 30 \, \mathrm{Mm}$. At that height, at about 4\,000 seconds we witness overarching coronal rain formation, as seen clearly with locally high density and low temperature. Locally perturbed density above the flux rope's top undergoes thermal instability as radiative cooling outweighs the background heating, increasing the density further while decreasing temperature with time until coronal rain is formed. The coronal rain then deviates away from our 1D vertical ray at $t=6\,000 \, \mathrm{s}$ as it moves and slides sideways along the magnetic arcades on which it formed downwards.  This shows that active additional heating at the bottom boundary, either due to an artificial heating source or magnetic reconnection, is not a necessary condition to form coronal rain. In our case, coronal rain forms above a rising flux rope due to the upwards total (gas, magnetic and ram) pressure during flux rope formation.

From the bottom panel of the same Fig.~\ref{fig:promform_1D_evolve}, another arc is present in the ram pressure panel beneath the one leading to the coronal rain. This ram pressure arc whose apex is at $\approx 20\, \mathrm{Mm}$ at $t=4\,500 \, \mathrm{s}$ also traces the formation of the prominence plasma, and this is situated within the flux rope that formed. When the flux rope is created, it scoops up the heavier density from the bottom and since it moves at a vertical speed of $v_y \approx 5 \, \mathrm{km} \, \mathrm{s}^{-1}$, its ram pressure becomes significant. Eventually, this scooped-up and pushed-up density undergoes thermal instability in the same manner as the coronal rain until a solar prominence is formed. The solar prominence matter then falls down due to the Rayleigh-Taylor instability which will be elaborated upon later. Interestingly,  the flux rope's centre increases slightly as time goes on, i.e. especially after 6\,500 seconds. This is a manifestation of mass escaping the flux rope. Since the prominence glides out the flux rope along its magnetic field lines (which can be seen in our accompanying movie), the gravitational force on the flux rope decreases and so the flux rope rises again until a new equilibrium has been found. 

Our solar prominence and the coronal rain are both high condensation end-products of thermal instability and possess the same characteristics. Their peak densities are in the order of  $10^{-13} \, \mathrm{g} \, \mathrm{cm}^{-3}$ and temperatures of $T \sim 0.01 \, \mathrm{MK}$. The value of the prominence density is in agreement with observations where typical prominence number densities $n$ lie in the range $10^{9} - 10^{11} \, \mathrm{cm}^{-3}$ or densities $\rho$ between $10^{-15} - 10^{-13} \, \mathrm{g} \, \mathrm{cm}^{-3}$ \citep{labrosse2010}.  \citet{li2022} have carried out simulations of coronal rain and they define coronal rain as plasma whose densities fulfil the criteria $n \geqslant 7\cdot10^{9} \, \mathrm{cm}^{-3}$ or $\rho \gtrsim 10^{-14} \, \mathrm{cm}^{-3}$ and temperature $T \leqslant 0.1 \, \mathrm{MK}$. Our coronal rain characteristics are in agreement with their findings: cold and dense rain blobs are present that reach temperatures in the order of $~10\,000$ K and densities of around $~10^{-13} \, \mathrm{g} \, \mathrm{cm}^{-3}$. Thus, both thermodynamic variables differ by two of magnitude with the ambient coronal domain which also agrees with the findings of \citet{fang2015}. Interestingly, the coldest temperature and largest density within the obtained coronal rainblobs have been detected to be $\approx 5 \, 200 \, \mathrm{K}$ and $\approx 6.3 \cdot 10^{-13} \, \mathrm{g} \, \mathrm{cm}^{-3}$ which correspond to a factor of 200 and 275 difference with the coronal environment, respectively.

 The morphology of our prominence differs significantly from the work of \citet{kaneko2017} despite that our simulation setup is similar in its overall facets.  Major differences between our simulations result from our vastly improved resolution, our more realistic radiative cooling curve $\Lambda(T)$ and our algorithmic strategy details, including our partially open bottom boundary. \citet{kaneko2017} use the radiative cooling curve of \citet{hildner1974} that has much less detailed variation with temperature than our \texttt{Colgan\_DM} radiative cooling curve. As this allows us to go down to prominence-relevant (order few 1000 K) temperature,  we can capture small-scale structures of the prominence.  As we will demonstrate later, increasing the resolution up to 20 km size cells gives rise to even more fine-scale structures. Hence,  the combination of a higher resolution and more refined radiative cooling curve is able to explain the discrepancy between our results and that of \citet{kaneko2017}. We will now analyze the full mass cycle in our high-resolution prominence realization.
 
\subsection*{Mass Cycle}
According to the mass continuity equation \eqref{eq:continuity},  the evolution of a system's total mass depends on the mass rates across the boundaries of the domain, i.e.  
\begin{equation}\label{eq:massrate}
\dfrac{\mathrm{d}M}{\mathrm{d}t} = \iiint_\mathcal{V} \dfrac{\mathrm{d}\rho}{\mathrm{d}t}  \mathrm{d}V = -\iiint_\mathcal{V} \nabla \cdot (\rho \mathbf{v}) \mathrm{d}V = -\oiint_S  (\rho \mathbf{v}) \cdot \mathbf{\hat{N}} \mathrm{d}S,
\end{equation}
where we have used the divergence theorem to rewrite the last equality, introducing the outward unit normal vector $\mathbf{\hat{N}}$ on all sides.  The \textit{mass draining rate} at a fixed height $y$ is computed by categorising those cells at the horizontal plane whose dot product -$\mathbf{v} \cdot \mathbf{\hat{N}}=v_y$ is negative and satisfies $T \leqslant 0.1 \, \mathrm{MK}$,  consistent with our working definition of  prominence matter. Since gravity is oriented in the negative $y$-direction, we have chosen the plane $y = 5 \,  \mathrm{Mm}$ to quantify a downward drainage rate, as this height is far enough from the bottom boundary.

\begin{figure*}[!t]
 \begin{center} 
\includegraphics[width=\textwidth]{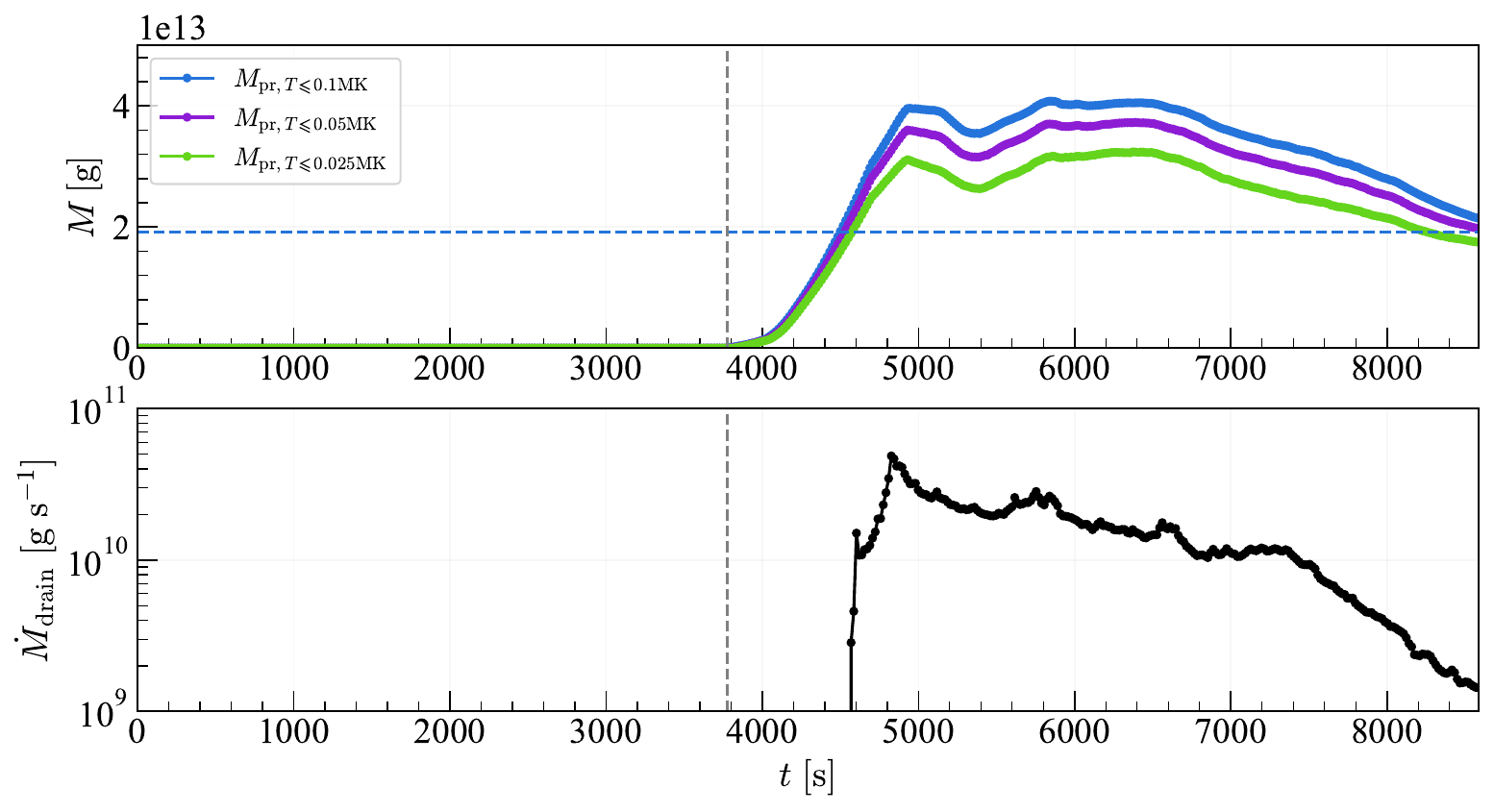}
 \caption{ Top panel displays the combined prominence-rain mass $M_\mathrm{pr}$ for three different temperature cut-off values.  The blue,  dashed,  horizontal line is an estimation of the mass reservoir within the flux rope.  Bottom panel shows the mass draining rate across a plane at height $y=5\,\mathrm{Mm}$. The vertical, dashed grey line exhibits the timestamp at which first condensations occur ($T \leqslant 0.1 \, \mathrm{MK}$). All quantities are shown in CGS-units. }\label{fig:mass}
 \end{center}
 \end{figure*}

If $\mathcal{V}(T\leqslant T_\mathrm{e}\, \mathrm{MK})$ specifies the regions where $T \leqslant T_\mathrm{e}$ MK, as visualized for $T_\mathrm{e}=0.1$ in panels C and D from Fig.~\ref{fig:promform},  then the combined mass of the prominence and coronal rain can be obtained by directly integrating the density over this volume at each time instant,  i.e.
\begin{equation}\label{eq:prom_mass}
M_{\mathrm{pr},T\leqslant T_\mathrm{e}} = \iiint_{\mathcal{V}(T\leqslant T_\mathrm{e} \mathrm{MK})}  \rho \mathrm{d}V \,.
\end{equation}
The prominence-rain mass $M_\mathrm{pr}$ and mass draining rate $\dot{M}_\mathrm{drain}$ are all acquired by numerically evaluating the above volume and surface integrals with the trapezoidal rule. Their time evolution can be seen in Fig.~\ref{fig:mass}.
 
 The top panel of Fig.~\ref{fig:mass} shows the mass of the condensations defined as $T \leqslant 0.1 \, \mathrm{MK}$ (blue curve),  $T \leqslant 0.05 \, \mathrm{MK}$ (purple curve) and $T \leqslant 0.025 \, \mathrm{MK}$ (green curve). The vertical, dashed grey line marks the time when the first condensations occur, i.e. some coronal plasma reaches temperatures of $T \leqslant 0.1 \, \mathrm{MK}$.  The horizontal,  blue dashed line is an estimate of the mass reservoir residing in the flux rope's interior which is equal to $1.9 \, \cdot 10^{13} \, \mathrm{g}$.  More information on how this mass reservoir is estimated can be found in Appendix A.  We obtain total condensation mass values in the order of $10^{13} \, \mathrm{g}$ which is in fair agreement with observed prominence masses \citep{labrosse2010}, though still on the lower side of observations. Note that most of this prominence-rain matter reaches temperatures below $T_\mathrm{e}=0.025$ MK, and has minimal value of about $5 \, 200 \, \mathrm{K}$. Using the two mass curves for $T \leqslant 0.1\, \mathrm{MK}$ (blue) and $T \leqslant 0.025\, \mathrm{MK}$ (green), we can estimate the percentage of which cold and dense condensations ($T \leqslant 0.025\, \mathrm{MK}$) constitute the total mass of the warmer condensations ($T \leqslant 0.1\, \mathrm{MK}$). At $t=5\,839 \, \mathrm{s}$, when the blue curve reaches its peak, the mass of the blue curve is equal to $M_{T\leqslant 0.1} = 4.1 \cdot 10^{13} \, \mathrm{g}$ and the contemporary mass of the green curve is $M_{T\leqslant 0.025} = 3.2 \cdot 10^{13} \, \mathrm{g}$, resulting in a mass percentage $M_{T\leqslant 0.025} / M_{T\leqslant 0.1} = 78\%$. At the end of the simulation, this mass percentage grows to $M_{T\leqslant 0.025} / M_{T\leqslant 0.1} = 82\%$. This indicates that our prominence is prone to evaporation and that most of the prominence mass is stored in very cold condensations, i.e. plasma with temperatures below $25\,000 \, \mathrm{K}$.

For the three different temperature cutoff definitions of condensations, it can be seen that the combined mass of prominence and coronal rain material is larger than the mass reservoir of the flux rope for the majority of the simulation time. Note that coronal rain does not lie within the flux rope, and most of the coronal rain has vanished by $t = 6\,950 \,  \mathrm{s}$ but the final prominence mass still lies well above the estimated mass reservoir of the flux rope.  This is a significant finding as it proves that the overall mass of the prominence cannot only be the mere product of in-situ condensations due to thermal instability.  An additional source of matter is required in order to explain the final mass values and as we will show later, this additional source of matter is realized by a continuous mass influx from below.  
 
The bottom panel provides the mass draining rate $\dot{M}_\mathrm{drain}$ of condensations ($T \leqslant 0.1 \, \mathrm{MK}$) passing through the plane $y=5 \, \mathrm{Mm}$. Note that this mass drainage evaluation is calculated independently from the total mass computation, so they are not related a priori. While the mass drainage rate is evaluated through the plane $y=5 \, \mathrm{Mm}$, the total mass is computed for the entire simulation domain $y\geqslant 0 \, \mathrm{Mm}$\footnote{Since the mass drainage rate depends on the velocity and the velocity is set to zero at the lower boundary, the mass drainage rate is chosen to be evaluated at $y=5 \, \mathrm{Mm}$ to minimise artefacts from the bottom boundary condition.}. The mass draining rate shows a strong peak at $t \approx 4\, 750\, \mathrm{s}$,  reaching a value of $\dot{M}_\mathrm{drain} \approx -6\cdot 10^{10} \, \mathrm{g} \, \mathrm{s}^{-1}$.  This peak originates from the coronal rain that travels at an average speed of $\vert \mathbf{v} \vert = 57 \, \mathrm{km} \, \mathrm{s}^{-1}$ when passing the reference plane.  The average vertical component of the velocity is $v_y = - 32  \, \mathrm{km} \, \mathrm{s}^{-1}$. The speeds of the condensations $\vert \mathbf{v} \vert$ range from $0.4 \, \mathrm{km} \, \mathrm{s}^{-1}$ to $159 \, \mathrm{km} \, \mathrm{s}^{-1}$.  The vertical velocities $v_y$ vary from  $-0.01 \, \mathrm{km} \, \mathrm{s}^{-1}$ to $-92.5 \, \mathrm{km} \, \mathrm{s}^{-1}$. The almost stationary condensations are due to the solar prominence and not the coronal rain. Note that the blob with the fastest vertical velocity does not necessarily correspond to the rainblob with the fastest speed.  These speeds have been observed for coronal rain (see Fig.  6 of  \citet{antolin2012}).  The mass draining rate then steadily declines as the coronal rain stops forming and the prominence mass is decreasing.
 
The average prominence growth rate $ \langle \dot{M}_\mathrm{prom} \rangle$ and mass drainage rate $\langle \dot{M}_\mathrm{drain} \rangle$ are evaluated as well. The average mass drainage rate $\langle \dot{M}_\mathrm{drain} \rangle$ is calculated by summing over all non-zero $\dot{M}_{\mathrm{drain}, i}$ datapoints and subsequently dividing by the total number of datapoints within this sample. For the average mass drainage rate, this implies that only datapoints for $t\geqslant 4 \, 568 \, \mathrm{s}$ are considered. For the average prominence growth rate $ \langle \dot{M}_\mathrm{prom} \rangle$, the gradient of $M_{\mathrm{pr},T\leqslant 0.1 \, \mathrm{MK}}$ is first evaluated for all $t$ and subsequently the same averaging process is conducted but for all non-zero $\dot{M}_{\mathrm{pr},T\leqslant 0.1 \, \mathrm{MK}, i}$ datapoints, i.e. for those datapoints that satisfy $t\geqslant 3 \,761 \, \mathrm{s}$. Note that in the averaging process of $\langle \dot{M}_\mathrm{prom} \rangle$ that both positive and negative gradients are present (see blue curve in top panel of Fig. \ref{fig:mass}); they have not been differentiated during the averaging process. The average rates are equal to $ \langle \dot{M}_\mathrm{drain} \rangle = -2.9 \cdot 10^{10}  \, \mathrm{g} \, \mathrm{s}^{-1}$ and  $ \langle \dot{M}_\mathrm{prom} \rangle = 4.4 \cdot 10^{9}  \, \mathrm{g} \, \mathrm{s}^{-1}$. Since the prominence mass rate is the net result of the condensation growth rate and the mass drainage rate 
\begin{equation}
\langle \dot{M}_\mathrm{prom} \rangle = \langle \dot{M}_\mathrm{cond} \rangle + \langle \dot{M}_\mathrm{drain} \rangle,
\end{equation}
the average condensation rate can be obtained, resulting in a value of $\langle \dot{M}_\mathrm{cond} \rangle =  3.3 \cdot 10^{10}  \, \mathrm{g} \, \mathrm{s}^{-1}$. These average mass rates are in good agreement with \citet{liu2012}. They have observed a draining quiescent prominence through \textit{SDO}/Atmospheric Imaging Assembly EUV channels and found that the quiescent prominence possesses a continual mass cycle of drainage and supply. Their average mass drainage rate is equal to $ \langle \dot{M}_\mathrm{drain} \rangle =  1.1 \cdot 10^{10} \, \mathrm{g} \, \mathrm{s}^{-1}$ and their average mass condensation rate is equal to $ \langle \dot{M}_\mathrm{cond} \rangle =  1.2 \cdot 10^{10} \, \mathrm{g} \, \mathrm{s}^{-1}$. It should be noted that coronal rain contributes to our mass rates whereas the mass rates from \citet{liu2012} are solely due to the solar prominence. Our findings agree very well, nevertheless, and in our results this mass drainage consists of plasma undergoing mRTI and plasma gliding along magnetic field lines towards the bottom, parallel to the spine of the solar prominence.  \citet{berger2012} observed a draining prominence that subsequently forms again.  During the disappearance of the observed prominence,  it can be seen through the 304 \AA \ SDO/AIA EUV filter that dense and cold plasma leaves the solar prominence parallel to the LOS (see their supplemented animation of their Fig.~1). From our results this mass escape occurs due to the plasma gliding along strongly sheared, asymmetric, dipped magnetic arcades (see also our further Fig.~\ref{fig:helicity}).

 \begin{figure*}[!t]
 \begin{center} 
 \includegraphics[width=\textwidth]{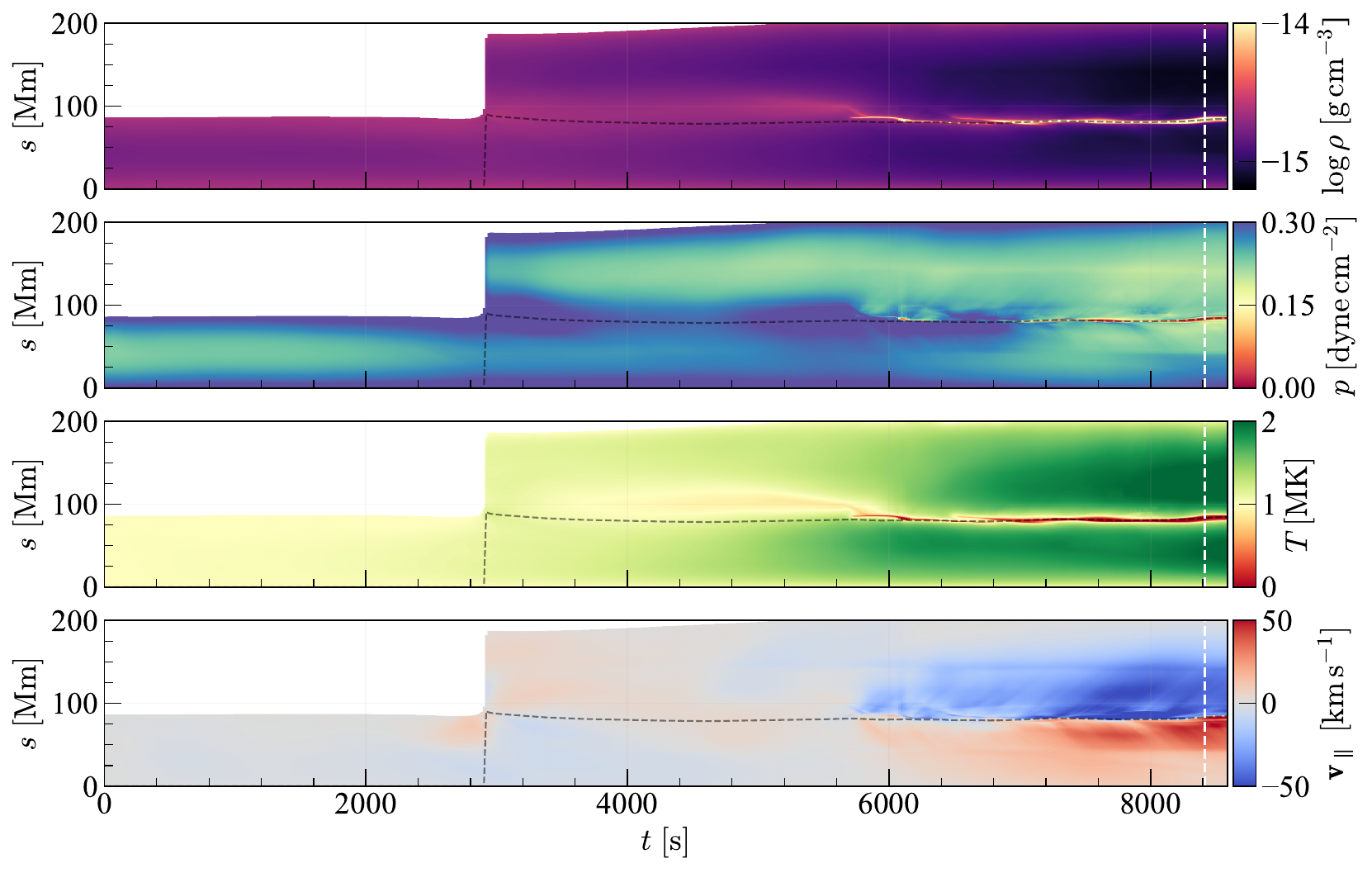}
 \caption{Temporal evolution of the logarithmic density $\log \, \rho$,  the pressure $p$,  the temperature $T$ and the tangent velocity $v_\parallel$ along a magnetic field line.  $s \ [\mathrm{Mm}]$ is the arclength of the magnetic field line. The black dashed line indicates the position of the dip within the magnetic field line. The white, vertical,  dashed line marks the time at which the 1D plot of Fig.~\ref{fig:sl_variables_1D} is taken. Colours are saturated to highlight contrasts. The non-coloured region in the left corner of the four panels is due to the fact that the magnetic field line is not reconnected yet.}\label{fig:sl_variables}
 \end{center}
 \end{figure*}

The mass condensation rate that we quantified as identical to observations is a contribution from in-situ condensations due to thermal instability and plasma siphoning from our lower boundary towards the prominence formation region. This mass siphoning and its physical cause are directly observed in Fig.~\ref{fig:sl_variables} which exhibits the temporal evolution of the density $\rho$,  the pressure $p$,  the temperature $T$ and the tangent velocity $v_\parallel$ along a representative, individual magnetic field line which develops prominence matter. The tangent velocity $v_\parallel$ is computed as $v_\parallel = \mathbf{v} \bullet \hat{\mathbf{T}}$ with $\hat{\mathbf{T}} = \mathbf{B}/B$. At $t=0 \, \mathrm{s}$ this magnetic field line is a magnetic arcade without any upwards dipped section, which has a length of $\approx 90 \, \mathrm{Mm}$. After $t=2\,900 \, \mathrm{s}$ the magnetic field line reconnects,  drastically increasing its length to $\approx 200 \, \mathrm{Mm}$ and taking on a dipped, helical configuration. The density $\rho$ is shown in the top panel and there it can be seen that the reconnected magnetic field line scoops up the denser plasma from the bottom regions which gradually condenses until a solar prominence is formed at $t\approx 5 \, 700 \, \mathrm{s}$.  Once the prominence is formed,  it sucks in surrounding plasma from further away regions on the magnetic field line,  increasing the density of the prominence within the dip to values above $10^{-14} \,  \mathrm{g} \, \mathrm{cm}^{-3}$ and decreasing the density from the surrounding to below $10^{-15} \,  \mathrm{g} \, \mathrm{cm}^{-3}$ for $s=50 \, \mathrm{Mm}$ and $s=125 \, \mathrm{Mm}$.  It can further be noted that the prominence does not form within the dip (this location is indicated by a black dashed line) but rather some distance away from it for $t <   5 \, 700 \, \mathrm{s}$. That the prominence sucks the surrounding plasma and that it forms off-dip arises from the interplay between the pressure,  tangent velocity and radiative cooling and aligned thermal conduction on the magnetic field lines.

That the prominence is not originally formed within the dip is explained by looking at the pressure distribution on the magnetic field line, which is provided by the second panel in Fig.~\ref{fig:sl_variables}. When the magnetic field line is reconnected,  a large pressure region within the dip at $t = 2\,900 \, \mathrm{s}$ is sandwiched between two lower pressure regions. Since the region $s > 90 \, \mathrm{Mm}$ for $t \geqslant 2 \, 700 \mathrm{s}$ has a lower pressure than the opposite side of the dip,  a stronger pressure gradient is present which pulls the plasma away from the dip. The condensing plasma then finds equilibrium at that position on the magnetic field line where the upwards pressure gradient is exactly counteracted by the downwards gravity and upwards Lorentz forces.  The plasma then maintains this equilibrium position until it condenses sufficiently enough for gravity to pull it back towards the dip.

Once the prominence is formed,  radiative cooling outbalances both the background heating and thermal conduction to maintain a cool plasma that also possesses an extremely low pressure of $0.02 \, \mathrm{dyne} \, \mathrm{cm}^{-2}$.  Due to this low pressure of the prominence,  a high pressure gradient arises that is directed towards the prominence plasma,  sucking all the plasma within the magnetic flux bundle around this field line towards the prominence.  This is in agreement with \citet{antiochos1991} who found that condensations induce pressure gradients as well.  This can explicitly be seen from the bottom panel of Fig.~\ref{fig:sl_variables} that shows the tangent velocity $v_\parallel$. Note that $v_\parallel$ is positive when it is pointed towards the same direction as $\hat{\mathbf{T}}$ (which points to the direction of increasing arclength $s$) and negative when both vectors are anti-parallel with respect to each other.  This large pressure gradient is significant as the plasma reaches velocities up to $80 \, \mathrm{km} \, \mathrm{s}^{-1}$.  

 \begin{figure*}[!t]
 \begin{center} 
 \includegraphics[width=\textwidth]{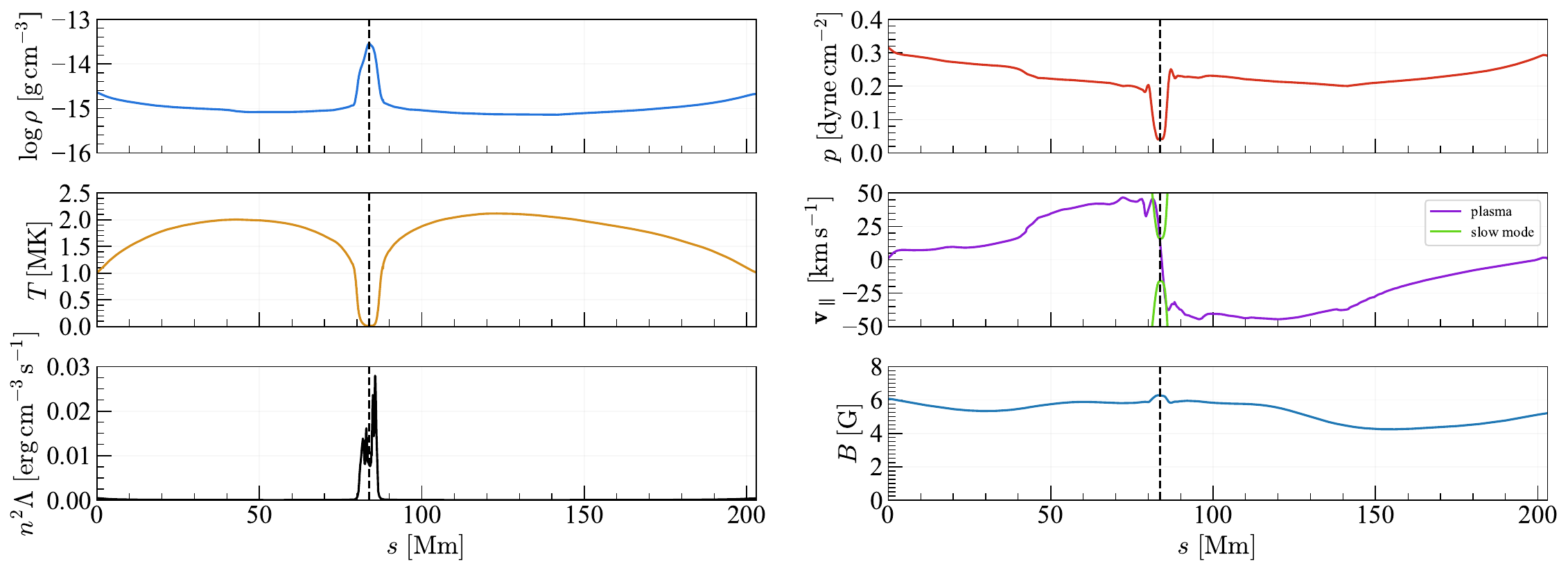}
 \caption{Variation of the logarithmic density $\log \, \rho$,  the pressure $p$, the temperature $T$, the tangent velocity $v_\parallel$, the net radiative losses $n^2 \Lambda$ and the magnetic field strength in their provided units along a representative reconnected magnetic field line at $t= 8 \, 416 \, \mathrm{s}$, indicated by the white dashed line in Fig. ~\ref{fig:sl_variables}.  The vertical,  black,  dashed line indicates the position of the prominence,  defined as the peak in density.  In the middle right panel,  the velocity of the plasma is compared with the speed of the slow wave mode.}\label{fig:sl_variables_1D}
 \end{center}
 \end{figure*}

Fig.~\ref{fig:sl_variables_1D}, taken at the time of the white dashed line in Fig.~\ref{fig:sl_variables},  shows how the variables vary along the magnetic field line at a constant time $t= 8 \, 416 \, \mathrm{s}$. The top left panel displays the logarithmic density and its peak is used to define the location of the solar prominence,  indicated as a black dashed line.  The pressure in the top right panel shows two sudden drops close to the prominence, inferring the presence of two stationary shockwaves. These two shockwaves are slow mode shockwaves, also referred to as \textit{rebound shocks}, since the tangent velocity (bottom right panel) there exceeds the speed of the slow wave mode.  These results are in agreement with literature as these rebound shocks have also been observed in 2D and 3D more idealized local numerical setups of solar condensation formations \citep{fang2013, fang2015, hermans2021, li2022} and 3D more idealized local numerical setups of solar condensation formations \citep{claes2020}. From the same panel, it can be seen that at the left footpoints of the magnetic field line,  i.e.  $s=0 \, \mathrm{Mm}$,  the tangent velocity is positive,  indicating that plasma is indeed being siphoned across the bottom boundary, consistent with our allowance for a partially open bottom boundary.

The temperature along the magnetic field line, middle left panel of the same Fig.~\ref{fig:sl_variables_1D},  shows that in the vicinity of the cold solar prominence, the surrounding temperature shoots to typical coronal values of about $1 - 2 \, \mathrm{MK}$.  More precisely,  the temperature of the prominence at $s\approx 83.8 \, \mathrm{Mm}$ is equal to $T = 0.01 \, \mathrm{MK}$ whereas the surrounding plasma with temperature $T=1 \, \mathrm{MK}$ is located at $s\approx79.4 \, \mathrm{Mm}$, resulting in an average temperature gradient of $\Delta T / \Delta s \approx 0.002 \, \mathrm{K} \, \mathrm{cm}^{-1}$. Because the dimensions of the cells around the prominence region are equal to 41.7 km, this jump is captured by 108 cells and hence is well-resolved. This region is identified as the prominence-corona transition region (PCTR). In the current case the length scale of the PCTR is equal to $4.5 \, \mathrm{Mm}$. However, the lengthscale of the PCTR varies through time and for different magnetic field lines. The smallest PCTR length scale is measured to be $708\, \mathrm{km}$ and is captured by 17 cells and hence is also well-resolved. 

That the prominence is able to maintain its temperature despite being in the vicinity of high temperature gradients indicates that radiative cooling outbalances the other heating terms significantly. The net radiative losses $n^2 \Lambda$ is provided by the bottom left panel and it shows a strong peak which does not coincide with the location of the prominence. The reason is due to the strong varying nature of the radiative cooling rate $\Lambda(T)$. At the density peak (black dashed line) the temperature is $T_1 \approx 11\, 000 \mathrm{K}$ and at the peak of the radiative loss $n^2 \Lambda$ the temperature is $T_2 \approx 51 \, 000 \, \mathrm{K}$. While the densities at these two temperatures differ by factor $\rho(T_2)/\rho(T_1) \approx 2$, the radiative cooling rate differs by a factor $\Lambda(T_2)/\Lambda(T_1) \approx 12$. Therefore, due to the fact that the densest and coldest condensation at $T=T_1$ radiates less energy than the warmer condensation at $T=T_2$, the peak of the radiative loss does not coincide with the density peak. The same graph also explains why the lengthscale of the PCTR is so small relative to the dimensions of the condensation. Within the PCTR, the radiative cooling rate is very sensitive to changes in temperature. To illustrate this, going from coronal temperatures $T_\mathrm{c} \sim 1 \, \mathrm{MK}$ to prominence temperatures $T_\mathrm{p} \sim 0.01\,\mathrm{MK}$, the net radiative cooling rates differ by a factor $\Lambda(T_\mathrm{p})/\Lambda(T_\mathrm{c}) \sim 10^{16}$! And as densities between the corona and prominence differ by two order of magnitude in addition, the net radiative losses $n^2\Lambda$ differ substantially within the PCTR whereas within the prominence itself the net radiative losses share the same order of magnitude as can be clearly seen from the bottom left panel.

The magnetic field strength along the fieldline is provided by the bottom right panel. It is evident that the magnetic field line is not symmetric. A second interesting feature is the little bump at the prominence location which is due to the prominence compressing the underlying magnetic field lines. The ratio of the magnetic field strength at the location $T=0.1 \, \mathrm{MK}$, denoted as $B(T=0.1 \,\mathrm{MK})$ and at the top of the bump $B_\mathrm{top}$ results in a relative difference of $(B_\mathrm{top} -  B(T=0.1 \,\mathrm{MK})) / B_\mathrm{top} \approx 0.03$. Therefore, across the prominence region the magnetic field strength varies insignificantly.

 \begin{figure*}[!t]
 \begin{center} 
 \includegraphics[width=\textwidth]{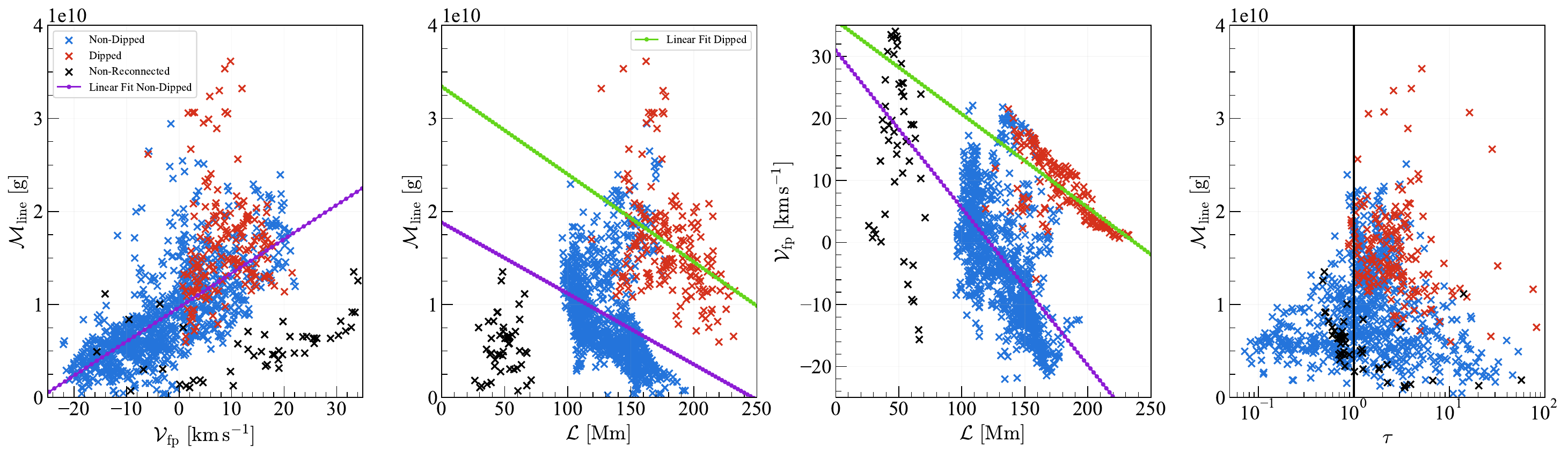}
 \caption{Statistical study of $1\,209$ magnetic field lines where the time-averaged prominence mass $\mathcal{M}_\mathrm{line}$,  the time-averaged flow velocity at the footpoints $\mathcal{V}_\mathrm{fp}$ and  the time-averaged line length $\mathcal{L}$ are plotted with respect to each other.  The relation $\mathcal{M}_\mathrm{line} \longleftrightarrow \mathcal{\tau}$ is examined separately. Note that the horizontal axis of the fourth panel is in logarithmic scale.  The data is split into three separate clusters: reconnected magnetic field lines that contain no dip (non-dipped blue crosses), reconnected magnetic field lines that have a dip (dipped red crosses) and magnetic field lines that have not reconnected (non-reconnected black crosses).  The purple solid line is a linear fit of the non-dipped group and the green line a linear fit of the dipped group.  Only groups with a correlation are linearly fitted.}\label{fig:pearson}
 \end{center}
 \end{figure*}

\subsubsection*{Statistical Analysis of Siphoning along Field Lines}
Up to now,  mass siphoning has only been discussed for an individual magnetic field line.  We will now quantify the importance of mass siphoning for the entire setup by defining the following three parameters for a very large sample of magnetic field lines: the instantaneous prominence mass along a magnetic field line $M_\mathrm{line}(t)$ at time $t$, the instantaneous net flow velocity $v_\mathrm{flow}(t)$ at the footpoints of this magnetic field line and the instantaneous length of the magnetic field line $L(t)$.  For an individual magnetic field at an instant time $t$,   its prominence mass is computed as 
\begin{equation}\label{eq:threadmass}
M_\mathrm{line}(t) = \sum_{\substack{i \\ T \leqslant 0.1 \, \mathrm{MK}}} \rho(s_i,  t) \cdot \pi R^2 \Delta s \,,
\end{equation}
with $s_i$ the local arc length value within the magnetic field line.  Here,  we treat the one dimensional line as a magnetic thread with radius $R$.  Hence, the sum is evaluated as a partition of smaller cylinders of volume  $\pi R^2 \Delta s$.  It is observationally agreed that threads in a solar prominence  have an average radius of $R = 200 \, \mathrm{km}$ \citep{lin2011}, so we adopt this constant value along all field lines (we could alternatively work with local flux conservation, but this simpler method agrees fairly well as we have shown previously that the magnetic field strength across the prominence region stays constant to a high degree).  Only those regions in the thread that have a temperature of $T \leqslant 0.1 \, \mathrm{MK}$ are evaluated in Eq.  \eqref{eq:threadmass}. The instantaneous, average flow velocity at the footpoints $v_\mathrm{fp}(t)$ is then calculated as
\begin{equation}
v_\mathrm{fp}(t) = \dfrac{1}{2n} \bigg(\sum_{\substack{i \\ s \leqslant 10 \, \mathrm{Mm}}} v_\parallel(s_i,  t) \ + \sum_{\substack{i \\s \geqslant (L(t) - 10 \, \mathrm{Mm})}} -v_\parallel(s_i,  t) \bigg) \,.
\end{equation}
Here, we average over a distance of $10 \, \mathrm{Mm}$ from each endpoint, and $n$ denotes the number of discrete arc elements within the regions $s \leqslant 10 \, \mathrm{Mm}$ and $s \geqslant L(t) - 10 \, \mathrm{Mm}$. Since magnetic field lines are traced at a constant resolution of $41.7 \, \mathrm{km}$, $n$ is the same for both regions and independent of time. The factor two accounts for the equal contribution from the two summation terms. It should be noted that although the magnetic field line is traced at a constant distance of 41.7 km, the effective resolution of the tracing depends on the dimension of the local cells which varies from 41.7 km (inside the prominence region) to 667 km (the outskirts of the domain). In the current definition,  a positive $v_\mathrm{fp}(t)$ indicates a net inflow of mass into the magnetic field line whereas a negative $v_\mathrm{fp}(t)$ implies a net outflow of plasma.

Together with $L(t)$  these three variables are in addition normalised over time.  Only those timestamps $t$ where the minimal temperature of the plasma reaches below $T \leqslant 0.1 \, \mathrm{MK}$ are included in the averaging process.  If $N$ denotes the amount of timestamps where the requirement $\min\{T\} \leqslant 0.1 \, \mathrm{MK}$ is met,  then for an individual magnetic field line the time-averaged quantities of the mass $\mathcal{M}_\mathrm{line}$ and flow velocity $\mathcal{V}_\mathrm{fp}$ are equal to
\begin{equation}
\mathcal{M}_\mathrm{line} = \dfrac{1}{N}\sum_{\substack{t \\ \min\{T(t)\} \leqslant 0.1 \, \mathrm{MK}}}  M_\mathrm{line}(t) \,,
\end{equation}
and 
\begin{equation}
\mathcal{V}_\mathrm{fp} = \dfrac{1}{N}\sum_{\substack{t \\ \min\{T(t)\} \leqslant 0.1 \, \mathrm{MK}}}  v_\mathrm{fp}(t) \,,
\end{equation}
respectively. The length of the magnetic field line $L(t)$ also does not stay constant,  hence the same averaging process takes place,  resulting in the time-averaged length $\mathcal{L}$
\begin{equation}
\mathcal{L} = \sum_{\substack{t \\ \min\{T(t)\} \leqslant 0.1 \, \mathrm{MK}}}  L(t) \,.
\end{equation}
Lastly, in order to investigate the importance of mass siphoning, the timescale of the siphoning flow $\tau_\mathrm{siphon}$ reaching the condensation site compared to the lifetime of the condensation $\tau_\mathrm{cond}$ within a magnetic field line has to be examined.  Let $\mathcal{V}_\mathrm{siphon,  1}$ and $\mathcal{L}_\mathrm{siphon,  1}$ be the average tangent velocity and prominence location along the magnetic field line between footpoint 1 and the prominence location, respectively.  Due to the fact that the condensation is often located off-centre in the magnetic field line,  we define the same parameters $\mathcal{V}_\mathrm{siphon,  2}$ and $\mathcal{L}_\mathrm{siphon,  2}$ but for the other side of the magnetic field line between footpoint 2 and the prominence site.  The timescale of the siphoning then depends on the footpoint

\begin{equation}
\tau_\mathrm{siphon,  1} = \dfrac{\mathcal{L}_\mathrm{siphon,  1}}{\mathcal{V}_\mathrm{siphon,  1}} \qquad \tau_\mathrm{siphon,  2} = \dfrac{\mathcal{L}_\mathrm{siphon,  2}}{\mathcal{V}_\mathrm{siphon,  2}}.
\end{equation}

The siphoning timescale of a magnetic field line is then defined as $\tau_\mathrm{siphon} = \min(\tau_\mathrm{siphon,  1}, \tau_\mathrm{siphon,  2})$.   Larger $\mathcal{V}_\mathrm{siphon}$ and smaller $\mathcal{L}_\mathrm{siphon}$,  i.e.  condensations that are located closer to footpoints,  result in smaller siphoning flow timescales as expected.  

The lifetime of a condensation $\tau_\mathrm{cond}$ is retrieved by counting the amount of timestamps that a magnetic field line contains plasma with temperatures below $T \leqslant 0.1 \, \mathrm{MK}$.

We now introduce the \textit{interaction parameter} $\tau$

\begin{equation}
\tau = \dfrac{\tau_\mathrm{siphon}}{\tau_\mathrm{cond}}.
\end{equation}
 
Note that it is a dimensionless quantity.  If $\tau > 1$,  then siphoning flow does not reach the prominence within the duration of our simulation and hence has not affected the condensation.  If $\tau < 1$,  then the siphoning flow is able to feed the prominence within the simulation time.  Therefore,  the interaction parameter $\tau$ can also be regarded as a weighing function that quantifies the contribution of in-situ thermal instability versus plasma siphoning to the final condensation mass.

A total of $2\,000$ magnetic field lines have been analysed and traced by distributing seedpoints randomly and uniformly in the plane sector $x > 0 \,  \mathrm{Mm},  y=0 \,  \mathrm{Mm}$ and $z>0 \,  \mathrm{Mm}$ at the moment the first condensations ($T \leqslant 0.1 \,  \mathrm{MK}$)  occur.  As the system evolves,  the same seedpoints are reused to track magnetic field lines. This resulted in $1\,209$ magnetic field lines where we could find cold condensations or about $61\%$ of all total magnetic field lines for which the three above parameters have been evaluated.  The relationship among these three variables for these $1\,209$ magnetic field lines are shown in Fig. ~\ref{fig:pearson}.  The data has been divided into three clusters: reconnected magnetic field lines that possess no dip (\textit{non-dipped}), reconnected magnetic field lines that contain a dip (\textit{dipped}) and magnetic field lines that have not reconnected (\textit{non-reconnected}), indicated by the blue, red and black colours,  respectively. Among the reconnected field lines, the total number of non-dipped magnetic field lines contribute to about $84\%$ of the data sample whereas the dipped magnetic field lines make up for $16\%$. The purple solid line is a linear fit of the non-dipped data and the green line a linear fit of the dipped field line data. The Pearson correlation coefficients $\mathrm{PCC}$ of the four relationships are provided by Table~\ref{table:pearson} with their corresponding standard deviation
\begin{equation}
\sigma = \sqrt{\dfrac{1 - \mathrm{PCC}^2}{N_\mathrm{lines} - 2}} \,.
\end{equation}
\begin{table*}[!t]
\caption{Pearson correlation coefficients $\mathrm{PCC}$ for the two cluster groups \textit{dipped} and \textit{non-dipped} amongst the time-averaged prominence mass $\mathcal{M}_\mathrm{line}$,  the time-averaged flow velocity at the footpoints $\mathcal{V}_\mathrm{fp}$ and  the time-averaged line length $\mathcal{L}$.  The relation $\mathcal{M}_\mathrm{line}$ and $\mathcal{\tau}$ is calculated separately.  The number of magnetic field lines within the sample $N_\mathrm{lines}$ is provided in the last column. See main text for more information. The non-reconnected cluster has not been considered in this statistical study.}\label{table:pearson}
\begin{center}
\begin{tabular}{ P{20mm} P{25mm}  P{25mm}  P{25mm} P{25mm} P{20mm}   } 
 \hline
 \hline
&  $\mathcal{M}_\mathrm{line} \longleftrightarrow \mathcal{V}_\mathrm{fp}$  &  $\mathcal{M}_\mathrm{line} \longleftrightarrow \mathcal{L}$ &  $\mathcal{V}_\mathrm{fp} \longleftrightarrow \mathcal{L}$ & $\mathcal{M}_\mathrm{line} \longleftrightarrow \mathcal{\tau}$ & $N_\mathrm{lines}$ \\
non-dipped     & $0.71  \pm 0.02$ & $-0.33 \pm 0.03$ & $-0.56 \pm 0.03$ & $0.10 \pm 0.03$ & $972$\\ 
dipped         & $0.07 \pm 0.07$  & $-0.39 \pm 0.07$ & $-0.70 \pm 0.05$ & $-0.40 \pm 0.07$ & $184$ \\ 
 \hline
\hline
\end{tabular}
\end{center}
\end{table*}

Well within the provided errors, all of the four relations for the two cluster groups non-dipped and dipped exhibit a (linear) correlation amongst each other,  except for the relation $\mathcal{M}_\mathrm{line} \longleftrightarrow \mathcal{V}_\mathrm{fp}$ of the dipped cluster group and $\mathcal{M}_\mathrm{line} \longleftrightarrow \mathcal{\tau}$ for the non-dipped group.  Since only magnetic field lines that contain cool matter are considered,   the presence of mass siphoning naturally explains why the three variables $\mathcal{M}_\mathrm{line}$, $\mathcal{V}_\mathrm{fp}$ and $\mathcal{L}$  are correlated.  When $\mathcal{V}_\mathrm{fp}$ is large and positive,  more plasma is able to feed the solar prominence through mass siphoning,  increasing the prominence mass $\mathcal{M}_\mathrm{line}$ and hence a positive correlation between $\mathcal{M}_\mathrm{line}$ and $\mathcal{V}_\mathrm{fp}$.  

  \begin{figure*}[!t]
 \begin{center}
    \includegraphics[width=\linewidth]{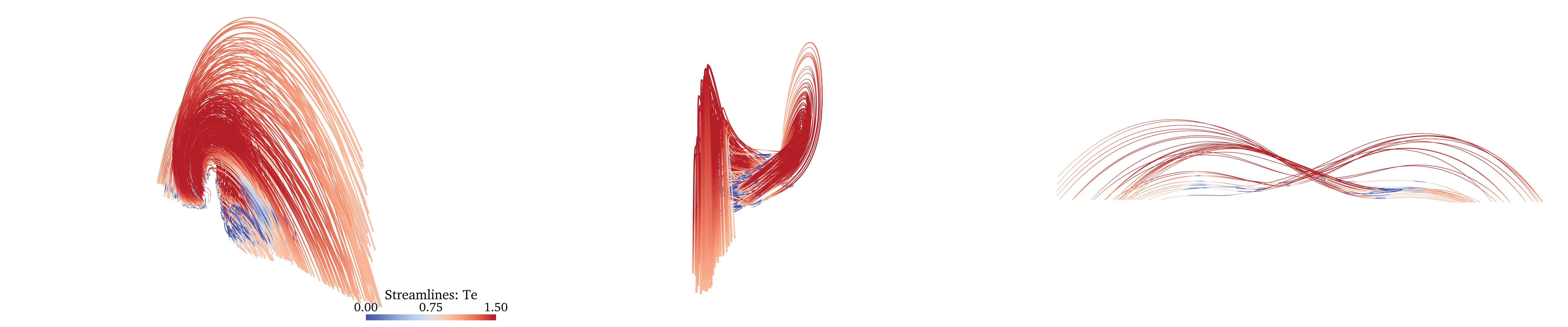}
 \end{center}
 \caption{Visualisation of all the magnetic field lines that belong to either the cluster group \textit{non-dipped} (left panel) or \textit{dipped} (middle panel) at a time when the prominence mass is well-developed. The right panel shows a subset of the dipped magnetic field lines from the middle panel where the maximum horizontal speed $v_\mathrm{hor} = (v_x^2 + v_z^2)^{1/2}$ of the condensations reaches above $70 \, \mathrm{km} \, \mathrm{s}^{-1}$.  The temperature $T \ [\mathrm{MK}]$ is plotted along the magnetic field lines.  The colours are saturated to highlight contrasts.  Different LOS angles are used between the three panels to accentuate their magnetic helicity.  The animation lasts for 12 seconds and starts from the moment when the first condensations occur ($T \leqslant 0.1 \, \mathrm{MK}$). It shows the growth of magnetic field lines with a registered condensation as well as the movement of the condensations along the magnetic field. (Animations of both figures are available in the online journal.)}\label{fig:helicity}
 \end{figure*}
 
$\mathcal{M}_\mathrm{line}$ and $\mathcal{L}$ are negatively correlated although a larger magnetic field line contains more plasma. However,  since mass siphoning arises due to large pressure gradients developing along the magnetic field line, the averaged strength of the pressure gradient decreases with increasing magnetic field line length. Remember that the low pressure of the solar prominence $p_\mathrm{prom}$ causes a large pressure gradient to exist and since pressure is strictly positive, it means that $\nabla p \sim (p_\mathrm{fp} - p_\mathrm{prom}) / \mathcal{L} \sim \mathcal{L}^{-1}$ with $p_\mathrm{fp}$ the (higher) pressure at the footpoints of the magnetic field line. This explains how a larger $\mathcal{L}$ results in a lower pressure gradient which in turn implies a lower  $\mathcal{V}_\mathrm{fp}$ and ultimately a lower $\mathcal{M}_\mathrm{line}$ since $\mathcal{M}_\mathrm{line}$ and $\mathcal{V}_\mathrm{fp}$  are positively correlated.  This line of reasoning is also confirmed by the right panel of Fig.  \ref{fig:pearson} which shows the negative correlation between $\mathcal{L}\leftrightarrow\mathcal{V}_\mathrm{fp}$.  Shorter magnetic field lines induce larger inflows $\mathcal{V}_\mathrm{fp}$ since $\mathcal{V}_\mathrm{fp} \sim \nabla p \sim \mathcal{L}^{-1}$ as we have shown.

Interestingly as well, there is a large discrepancy in the correlation value $\mathrm{PCC}$ of the relation $\mathcal{M}_\mathrm{line} \longleftrightarrow \mathcal{V}_\mathrm{fp}$ between the two groups,  well outside the range for their corresponding errors. In fact,  for the dipped group, we are unable to conclude whether a correlation even exists at all for these two variables.  This is due to the fact that the number of magnetic field lines $N_\mathrm{lines}$ within the dipped cluster group is much lower than the non-dipped group, resulting in that the corresponding errors on the $\mathrm{PCC}$ is larger as well.  The fourth panel shows that for almost all dipped magnetic field lines that $\tau > 1$. For the majority of the prominence magnetic field lines,  the siphoning flow does not reach the solar prominence whereas for non-dipped arcades  $\tau$ occurs at various values depending on the magnetic field line.  Despite that dipped magnetic field lines have a slightly higher siphoning velocity $\mathcal{V}_\mathrm{siphon}$ on average ($30 \, \mathrm{km} \, \mathrm{s}^{-1}$) than non-dipped magnetic field lines ($25 \, \mathrm{km} \, \mathrm{s}^{-1}$) ,  the prominence is located further away from the footpoints than for the non-dipped lines by a factor three on average. Since the \texttt{PCC} for the relation $\mathcal{M}_\mathrm{line} \longleftrightarrow\tau$ of the dipped magnetic field lines shows a negative correlation, a larger $\tau$ results in smaller $\mathcal{M}_\mathrm{line}$.  Therefore,  since for almost all of the prominence magnetic field lines there is no interaction between the prominence and siphoning flow,  no correlation exists between $\mathcal{M}_\mathrm{line}$ and $\mathcal{V}_\mathrm{fp}$. Hence, most of our simulated solar prominence is formed mainly due to in-situ condensations. 

The non-reconnected magnetic field lines that possess a condensation in Fig.  \ref{fig:pearson} are distinguished from the other two cluster groups for having a time-averaged line length $\mathcal{L} < 80 \, \mathrm{Mm}$, which can clearly be seen from the second and third panel.  The literature agrees that condensations cannot occur for those magnetic field lines whose length is smaller than the Field length since then thermal conduction would dominate over radiative cooling \citep{field1965,  smith1977,  kaneko2017}.  Hence,  rather than concluding that these magnetic field lines have formed a solar prominence or contain rain matter, it is much more likely that these magnetic field lines lie underneath the solar prominence and receive their condensations from mass slippage.

These results all confirm that mass siphoning is present in our simulation without any active prescription of evaporation or reconnection at the boundary: these siphoning flows are solely driven by thermal instability.  This has also been noted by \citet{xia2014} who have simulated a 3D solar prominence with an included chromospheric layer and obtained siphoning flows due to thermal instability. Noting that -- in contrast -- our simulation does not yet include the chromosphere with its transition region, it shows how mass siphoning from large coronal regions plays a role in prominence formation. We speculate that including the chromospheric regions will allow larger density plasmas to be siphoned to the prominence and perhaps allow more realistic prominence masses to be obtained in the order of $\sim 10^{14} - 10^{15} \, \mathrm{g}$.

In addition, the evaporation-condensation model strictly assumes that the heating length scale of the implemented localised heating source should be small (hence localised) in order not to heat up the prominence. Our results show that an ad-hoc additional localised heating term at the footpoints of the magnetic field line is not a necessary condition in order to promote siphoning flows or condensations. The first and third panel of Fig.  \ref{fig:pearson} show that the speed of these siphoning flows come in the range of  $0 - 35 \, \mathrm{km} \, \mathrm{s}^{-1}$.  It should be noted that our partially open bottom boundary limits the strength of the inflow since $v_y = 0$ is assigned to the cell-centers of the ghostcells below the bottom boundary. From Fig.~\ref{fig:sl_variables_1D} we see that a solar prominence is still able to survive,  despite that the temperature in its close surrounding is in the order of $1\, \mathrm{MK}$.

Finally, our 3D pure coronal volume simulation should be contrasted with the countless 1D hydro simulations where parametric prescriptions of additional, localized footpoint heating at transition region heights are a recurring ingredient \citep{xia2011, johnston2019, pelouze2022}. There, various authors argue the importance of Thermal-Non-Equilibrium (TNE) cycles that can only establish along loops of certain length and shape, under suitable heating (and chromospheric evaporation) prescriptions \citep{froment2018, klimchuck2019, antolin2022b}. These introduce additional arguments involving timescales of thermal and enthalpy fluxes into and out of the coronal parts of essentially line-tied flux ropes of given shape. Our setup deliberately excludes this TNE process by construction, as we do not have an underlying chromosphere and transition region, and only focus on the coronal evolution that follows when thermal instability sets in. We show explicitly that TNE is not a necessary condition for stimulating the development of condensations.

\subsubsection*{Magnetic Topologies of the Dipped- and Non-Dipped Cluster Groups}

Even though the correlation values $\mathrm{PCC}$ for the two relationships $\mathcal{M}_\mathrm{line} \longleftrightarrow \mathcal{L}$ and  $\mathcal{V}_\mathrm{fp} \longleftrightarrow \mathcal{L}$ agree fairly well between the two cluster groups \textit{dipped} and \textit{non-dipped} (w.r.t. their errors), Fig.~\ref{fig:pearson} shows that they exhibit noticeable differences. The major difference between the dipped- and non-dipped cluster groups is that the non-dipped group corresponds to arcade-like structures whereas the dipped group consists of only helical magnetic field lines. Both groups can be viewed in Fig. ~\ref{fig:helicity} at the time when the solar prominence is well-developed. This figure shows non-dipped (left) and dipped (middle) field lines that contain cool matter at this instant of time. From the animation, it is clear that as the simulation progresses, condensations spread over adjacent field lines for both cluster groups. This is a strong argument for thermal instability having the intrinsic tendency to spread over adjacent magnetic field lines. Earlier 2D coronal rain simulations \citep{fang2013} argued for fast magnetosonic perturbations that communicate sudden thermal changes across field lines.

The right panel of the same figure shows a subset of the dipped cluster group; those dipped magnetic field lines whose maximum horizontal velocity of the inhabiting condensation reaches above $v_\mathrm{hor} = (v_x^2 + v_z^2)^{1/2} \gtrsim 70 \, \mathrm{km} \, \mathrm{s}^{-1}$. These magnetic field lines represent some of the field lines whose condensations drain from the flux rope and glide towards the bottom of the simulation domain. It should be noted that the cutoff speed of $70 \, \mathrm{km} \, \mathrm{s}^{-1}$ has only been chosen to filter out many field lines in order to obtain a clear representation of how some condensations can escape the flux rope.  Although these magnetic field lines contain a dip, the dip is located close to the footpoints and thus these magnetic field lines have a low magnetic twist compared to those fieldlines whose dip lie roughly in the centre. Dipped magnetic field lines whose condensations reach high horizontal velocities and are able to escape the flux rope manifest a strongly sheared and asymmetric, dipped arcade-like character. Magnetic field lines with low horizontal velocities are also those where the dip lies more central and have more twist.

In our simulation, we thus retrieve all relevant magnetic field line topologies in the context of coronal condensations, i.e. non-dipped arcades, sheared and dipped arcades and helical flux ropes. Indeed, only few works reconcile flux ropes and (sheared) arcades in the magnetic configuration of solar filaments \citep{guo2010}. From our results, the three mentioned magnetic topologies are all essential for solar prominences and each carry a distinct, important role: non-dipped magnetic arcades are associated with coronal rain, helical magnetic field lines allow for the prominence to be nested in the flux rope and strongly sheared arcades help condensations to escape the flux rope. These all play a role in the mass budget of the filament.

\subsection*{Magnetic Rayleigh-Taylor Instability}
\begin{figure*}[!t]
 \begin{center}
    \includegraphics[width=\linewidth]{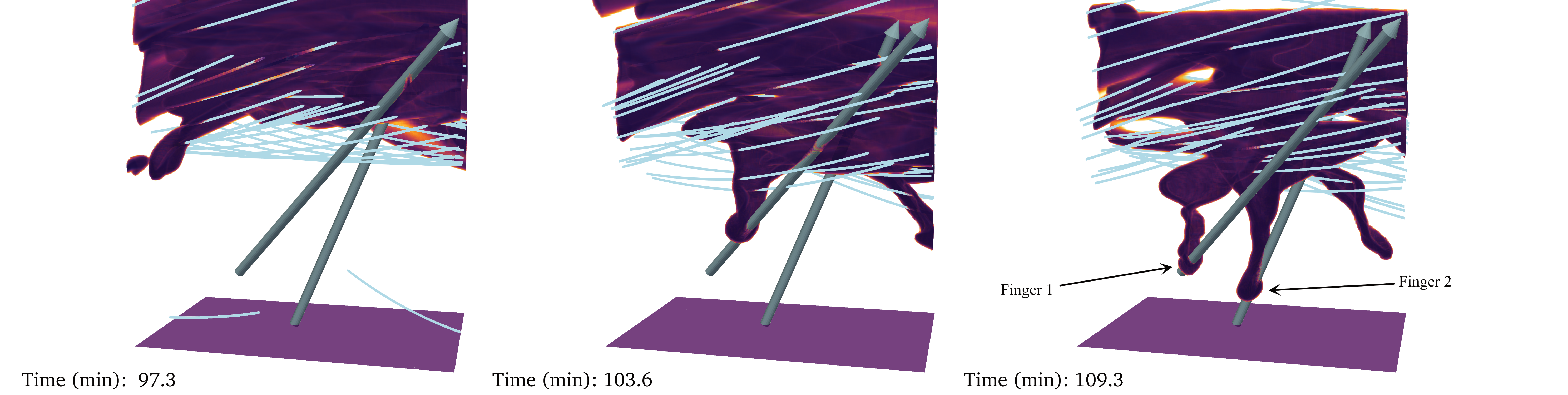}
 \end{center}
 \caption{Evolution of the mRTI instability. The solid lines are magnetic field lines and the volume rendering shows those cells whose temperature $T \, [\mathrm{MK}]$ lies below the cutoff value $0.1 \mathrm{MK}$. The two arrows represent rays along which two falling fingers are studied, with Finger 1 and 2 as indicated at right. The unicolour bottom plane is only added to provide perspective to the system. The visualisation is constrained to $-3 \, \mathrm{Mm} \leqslant x \leqslant 3 \, \mathrm{Mm}$, $ y \leqslant 15 \, \mathrm{Mm}$ and $-5 \, \mathrm{Mm} \leqslant z \leqslant 5 \, \mathrm{Mm}$.  The animation lasts for 7 seconds and shows how the mRTI fingers form and evolve. (An animation of this figure is available in the online journal.)}\label{fig:mrti}
 \end{figure*}

Especially in the centre of the flux rope it is evident that the prominence does not remain static and shows clear fingering structures due to magnetic Rayleigh-Taylor instability (mRTI).  Fig.  \ref{fig:mrti} shows three snapshots of the evolution of three distinct mRTI-fingers,  an animation of which is available in the online journal.   The fingers have a width of $1 - 1.5 \,  \mathrm{Mm}$ which is also in fair agreement with the widths of vertical structures in observed solar prominences (see Fig.~5 of \citet{jenkins2022}).
 
\begin{figure*}[!t]
 \begin{center} 
 \includegraphics[width=\textwidth]{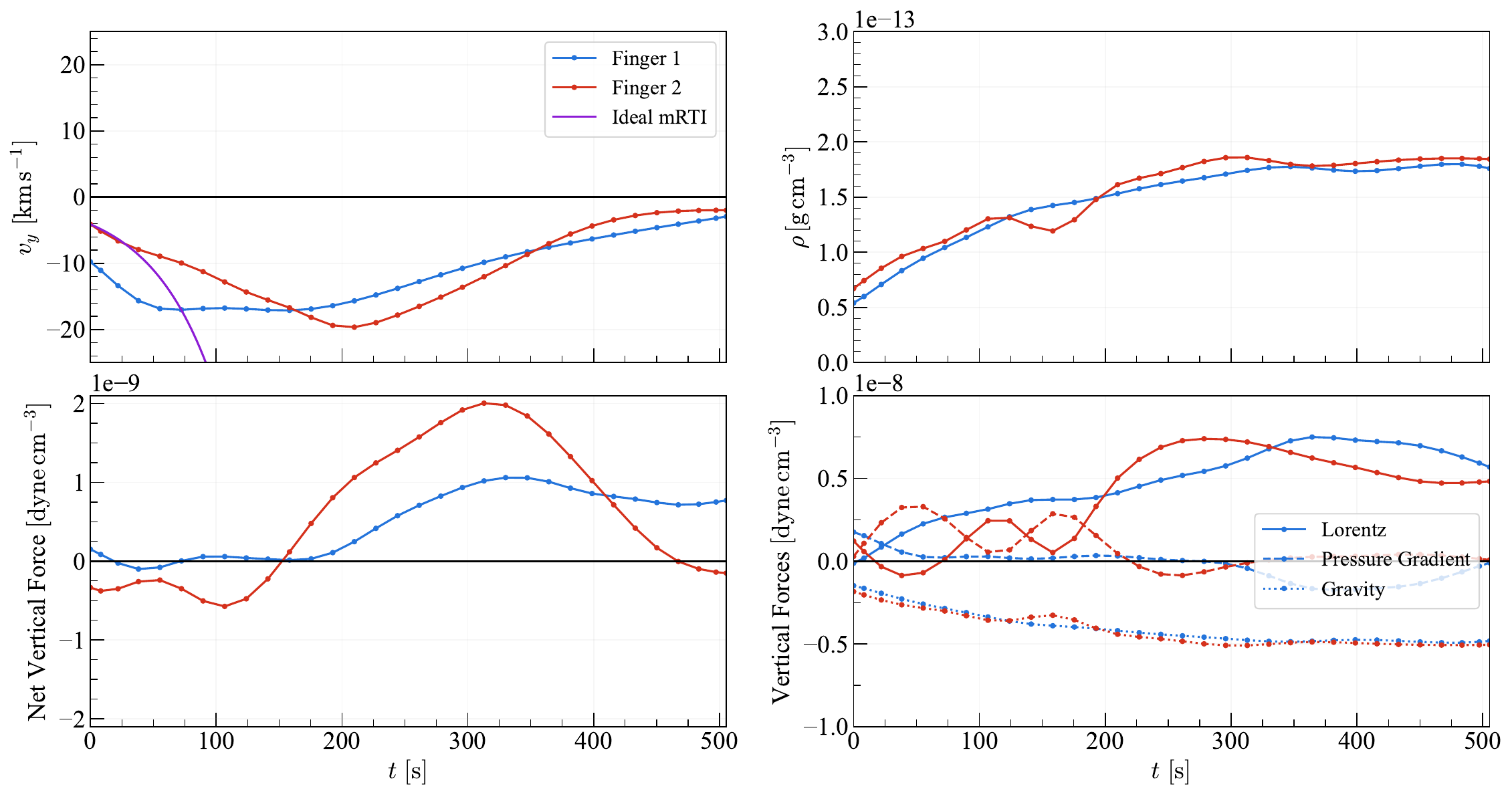}
 \caption{The evolution of the vertical velocity $v_y$ provided with the analytical solution of the linear mRTI in purple (top left panel),  the density $\rho$ (top right panel),  the net vertical force (lower left panel) and the vertical components of gravity,  the pressure gradient and the Lorentz force (lower right panel) are shown along the two rays in Fig.~\ref{fig:mrti} in function of time $t$.  Blue-coloured curves refer to the dynamics of Finger 1 and red-coloured curves to the dynamics of Finger 2. The black, solid line in the bottom two panels is to signify the zero level. In the lower right panel, the Lorentz force is indicated by the solid curve,  the pressure gradient by the dashed curve and gravity by the dotted curve for the two fingers. Note that the time $t$ does not mean the global time of the simulation, but rather the elapsed time after both fingers started falling along their analyzed rays.}\label{fig:mrti_1D}
 \end{center}
 \end{figure*}

In the movie, it can be seen that the left-most finger (Finger 1) shoots out of the flux rope and decelerates strongly until it is halted mid-air whereas the middle finger (Finger 2) accelerates towards the bottom. Both fingers display unique dynamics and in order to quantify their properties, two rays are chosen to analyse their trajectories. We examine the evolution of the density $\rho$,  vertical velocity $v_y$ and the vertical forces $F_y$ along the rays that essentially trace out the path of their tips.  The evolution of these parameters are in Fig.  \ref{fig:mrti_1D} in function of their elapsed time as they progress along their respective rays. We have applied a Gaussian kernel in order to filter out the noise from the data.
 
In the top left panel,  the vertical velocity $v_y$ of both fingers is shown in addition to the analytical solution of the linear mRTI theory for finger 2. Since our system is not the textbook mRTI setup (typically two uniform density states separated by a sharp interface), deviation from linear theory predictions can easily arise, and be due to actual non-linearity. Nevertheless, we notice that the evolution of finger 2 (red curve) follows the trend of the analytical, idealised mRTI solution (purple curve) very well for the first thirty seconds. After thirty seconds, non-linear effects start to become important and dominate the evolution of the mRTI. Finger 1 (blue curve) enters its ray with a higher speed at 10 km/s than finger 2 with a speed of 2 km/s, inferring that finger 1 is more evolved at the moment when it is detected.  However, it stops with accelerating after 50 seconds, reaching a maximum speed of $\approx 17 \,  \mathrm{km} \,  \mathrm{s}^{-1}$ whereas finger 2 continues to gain speed until at $\approx 20 \,  \mathrm{km} \,  \mathrm{s}^{-1}$. This can be seen from the bottom left panel which displays the net vertical force acting on the fingers. Both fingers go through different phases. For finger 1 after being decelerated there is a near-constant speed phase and subsequently a deceleration phase.  For finger 2, it goes through a strong acceleration phase,  a subsequent deceleration phase and ultimately a phase near zero velocity. The acceleration of finger 2 is a result of a smaller Lorentz force compared to finger 1 for $t \leqslant 100 \,  \mathrm{s}$. Since they have almost the same density (top right panel), their gravitational force can be approximated as near identical. After $t \geqslant 200 \,  \mathrm{s}$,  both fingers lose vertical speed $v_y$ since the net vertical force is positive. For the entire evolution, the densities of the two fingers are quasi-similar (top left panel), hence any differences should originate from either the Lorentz force or the pressure gradient. It can be seen that the Lorentz force is generally much greater than the pressure gradient in the lower right panel. In both fingers, the magnetic tension force is strictly positive whereas the magnetic pressure is strictly negative.  Hence,  due to the fact that the full Lorentz force is always positive, it implies that magnetic tension is the main decelerator and gravity and magnetic pressure the main accelerators. The pressure gradient switches between being an accelerator and decelerator as it occasionally turns positive and negative. As the system further evolves, all the forces on finger 2 balance each other out and result in an almost zero vertical velocity (top left panel).
 
  \begin{figure*}[!t]
 \begin{center} 
 \includegraphics[width=\textwidth]{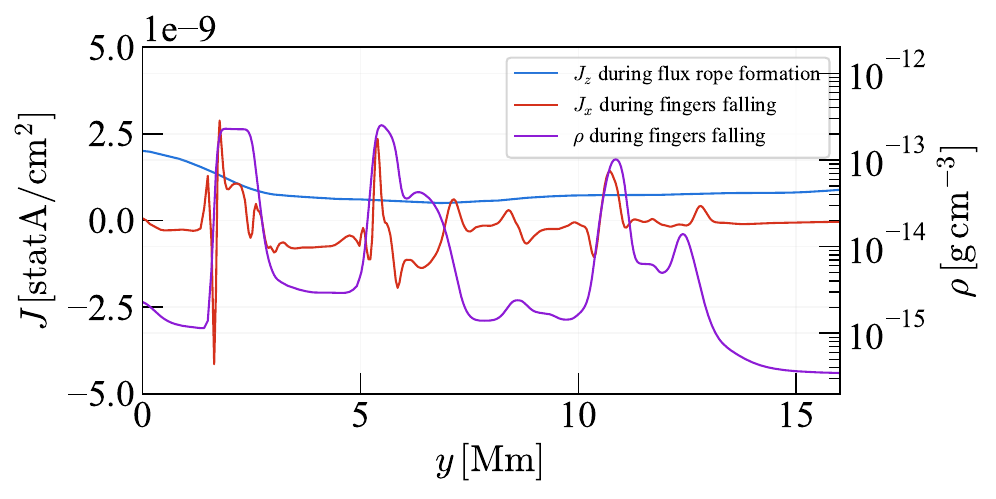}
 \caption{Spatial variation along the central $y$-axis ($x=z=0 \, \mathrm{Mm}$) of the $z$-component of the current density $J_z$ during the formation of the flux rope, and the $x$-component $J_x$ and the density $\rho$ during the evolution of the mRTI fingers. These $J_x$ and $J_z$ components are chosen since they manifest the strongest gradients at these respective times along the $y$-axis.}\label{fig:current_densities}
 \end{center}
 \end{figure*}
 
From the animation it can be noted that the mRTI fingers and the magnetic field lines are not always strictly tied together. The 3D ideal mRTI acts by interchanging field lines which thus in effect displace heavier material downwards. In our numerical study, we thus obtain some degree of cross-field mass transport or mass slippage for short \citep{low2012}. Mass slippage is also an actual resistive MHD phenomenon and arises in locations where gradients in the current density are large. From the induction Eq.~\eqref{eq:induction} this implies that a region is created where plasma diffuses across magnetic field lines since the source term $\nabla \times (\eta \mathbf{j})$ becomes significant.  Although the anomalous resistivity prescription is set to zero well before the mRTI fingers appear, there will always be some numerical resistivity $\eta_\mathrm{num}$ present due to discretisation errors of the conservation equations. In solar prominences, the presence of neutrals can introduce additional effects such as ambipolar diffusion. This causes e.g. fast waves in the chromosphere to damp effectively when travelling across the magnetic field \citep{braileanu2021}. Therefore, a non-zero diffusion will be closer to reality than our ideal MHD run, and its effect on prominence dynamics still needs further study. Due to our high effective resolution, we expect a low numerical resistivity, but large gradient in the current density can compensate for a lower $\eta_\mathrm{num}$ value and hence make the source term $\nabla \times (\eta_\mathrm{num} \,  \mathbf{j})$ significant,  nevertheless. Fig.~\ref{fig:current_densities} shows $\rho$ and $J_x$ at the moment the fingers are falling and $J_z$ during the flux rope formation along the central $y$-axis ($x=z=0 \, \mathrm{Mm}$).  These components $J_x$ and $J_z$ exhibit the largest gradient at this respective time along the central $y$-axis and thus these components will also be the dominant components of $\nabla \times \mathbf{J}$.  During the flux rope formation,  $J_z$ shows minimal variation but magnetic reconnection is still able to manifest due to a large anomalous resistivity being active. When the mRTI fingers are formed and fall downwards,  it can immediately be seen that the $J_x$ shows much sharper gradients at the fronts of the mRTI fingers where the density suddenly jumps up to one or two orders of magnitude,  i.e.  to values $\rho \geqslant 10^{-13} \,  \mathrm{g} \, \mathrm{cm}^{-3}$.  This is not surprising since the heavier plasma of the mRTI fingers squeezes the underlying magnetic field lines together,  resulting in a sharp gradient in the magnetic field and hence in the current density as well.  Therefore,  the source term $\nabla \times (\eta_\mathrm{num} \,  \mathbf{j})$ becomes important underneath the mRTI fingers,  allowing for mass slippage to occur in addition to the mRTI. This has also been observed in a 2D setting of the levitation-condensation model of a simulated solar prominence \citep{jenkins2021}. The authors assigned a constant and uniform resistivity over the entire simulation domain and at an extreme resolution of 6 km, mass slippage occurred across nested field lines as well. 
 
  \begin{figure*}[!t]
 \begin{center}
    \includegraphics[width=\linewidth]{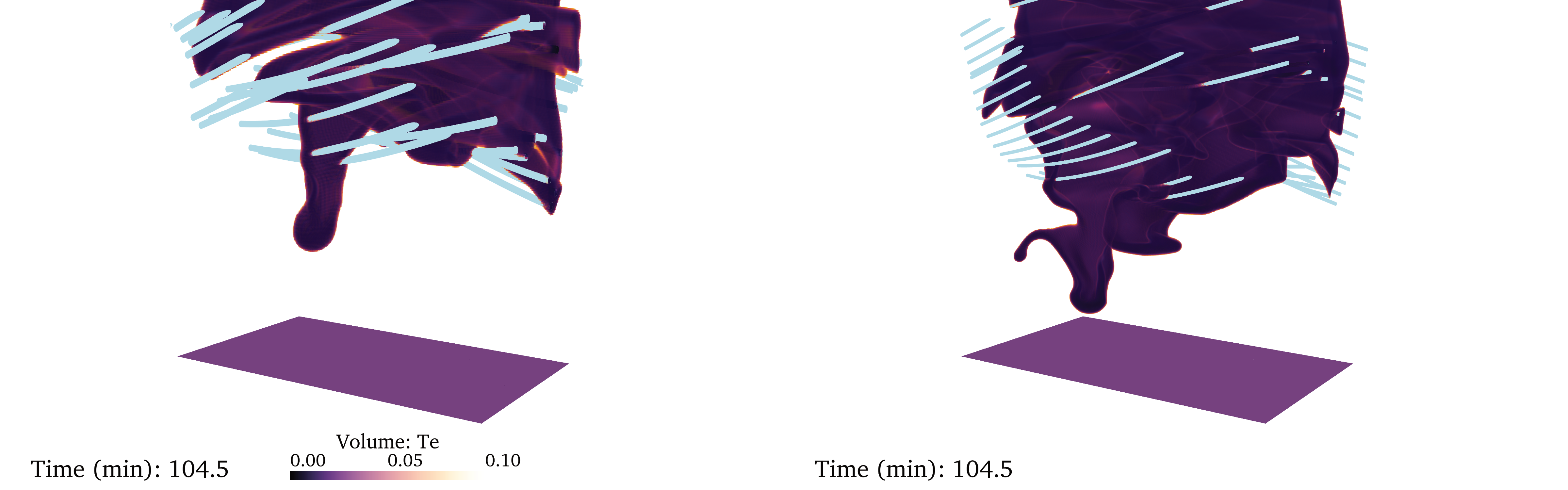}
 \end{center}
 \caption{The morphology of the mRTI fingers at the instantaneous time $t= 6\, 269 \,  \mathrm{s}$. The left panel shows the mRTI when the number of refinement levels is equal to four with an effective resolution of 41.7 km whereas the right panel is refined at five levels and capturs details at 20.8 km. The volume rendering displays the temperature $T \ [\mathrm{MK}]$.  The bottom plane at $y=0 \,  \mathrm{Mm}$ is added to bring a three-dimensional perspective to the volume. The visualisation is constrained to $-3 \, \mathrm{Mm} \leqslant x \leqslant 3 \, \mathrm{Mm}$, $ y \leqslant 15 \, \mathrm{Mm}$ and $-5 \, \mathrm{Mm} \leqslant z \leqslant 5 \, \mathrm{Mm}$.} \label{fig:highres}
 \end{figure*}
 
In order to assess the effects of resolution on the mRTI,  another simulation has been carried out where we reuse the simulation output of $t=5 \, 839 \, \mathrm{s}$ and increase the resolution onwards by augmenting the refinement level from four to five.  The effective resolution then sharpens from 41.7 km to 20.8 km.  However, increasing the refinement level by merely one level will prolong the computation time drastically.   Hence,  this increase of resolution is only applied within the column where the fingers are manifest,  i.e.   $\vert x \vert \leqslant 3 \,  \mathrm{Mm}$,  $y\leqslant 15 \,  \mathrm{Mm}$ and $\vert z \vert \leqslant 6 \,  \mathrm{Mm}$. Fig.  \ref{fig:highres} shows a visual comparison of the fingers that form when we employ four and five levels.  The increase of resolution affects the morphology significantly, giving rise to even finer structures. The dimensions of the little finger that pops out of the larger finger in the right panel is approximately $\Delta x \approx 0.26 \,  \mathrm{Mm}$ and $\Delta z \approx 0.40 \,  \mathrm{Mm}$.  It is not surprising that even finer-scale fingers develop in higher resolution runs.  Since the formation of the first finger in both the low- and high resolution simulations happen at the same time, we conclude that the most unstable length scale and growth rate of the mRTI is approximately the same for both cases. 

\subsection*{Synthetic Data}
 
   \begin{figure*}[!t]
 \begin{center}
    \includegraphics[width=\linewidth]{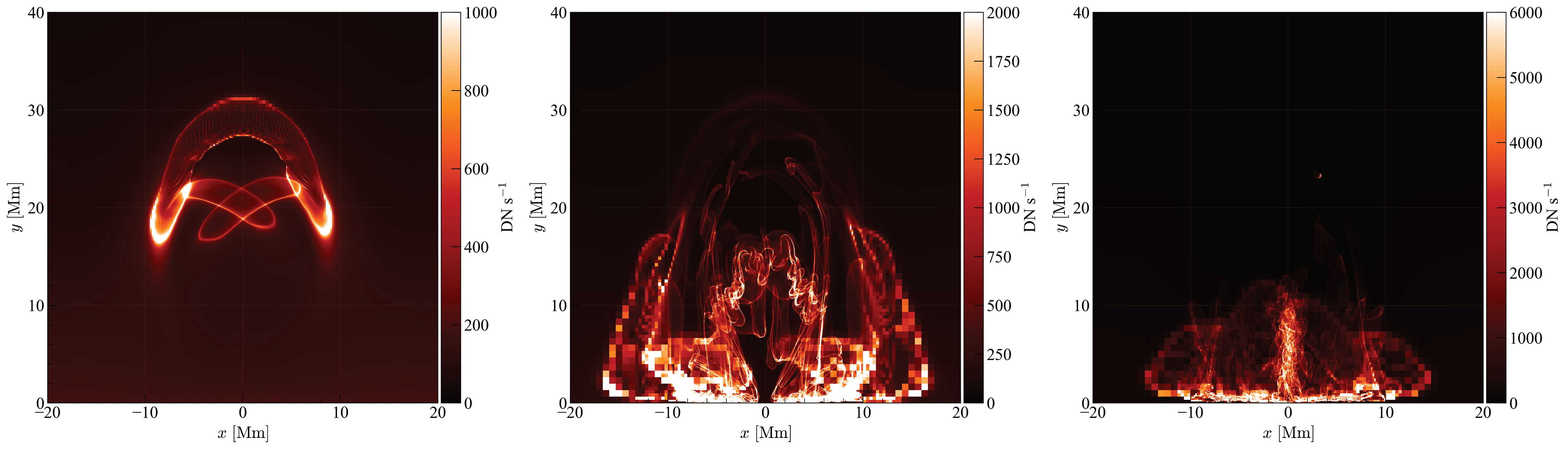}
 \end{center}
 \caption{Synthetic images of the spine of the solar prominence as seen through the 304 \AA \ SDO/AIA EUV filters.  The left panel is captured at the onset of the prominence- and coronal rain formation,   the middle panel during the falling towards the magnetic dips and the last panel when the mRTI fingers are manifested and distinct. At that final time, the material has centrally collected in the dips where the fingers also form. Colours represent the datanumber per second.  The animation lasts for 21 seconds and exhibits the spontaneous formation and falling of the condensations towards the dip.   (An animation of this figure is available in the online journal.)} \label{fig:304_spine}
 \end{figure*}
 
 \subsubsection*{304 \AA\ EUV Channel}
 In order to compare our results with observations, our numerical model is synthesized to the SDO/AIA EUV filters with wavelengths 193 \AA\, and 304 \AA\ \citep{sdoaia}.  Cold condensations appear as dark regions in 193  \AA \ since they don't emit light in this wavelength whereas in 304 \AA \ there is light emission when viewed off disk. Fig.~\ref{fig:304_spine} shows the synthetic views of our prominence at three distinct timestamps for the 304 \AA\ channel: the onset of formation,  the falling towards the magnetic dips and the moment when the mRTI fingers are manifested.  The line-of-sight (LOS) is taken parallel to the spine of the flux rope. In the left panel,  the initial formation of the coronal rain and solar prominence are evident. The solar prominence does not form within the magnetic dips as mentioned earlier and hence commences to fall while being guided by the magnetic field lines: this is evident in the middle panel. More regions (within and outside the flux rope) undergo thermal instability, resulting in the evolution of more condensed mass. The pixelated area in the middle panel is an artefact from our derefinement strategy as explained in Sect. 2: condensations that are located further from the centre are captured at the lowest refinement level.  Within these pixelated areas, there is a much higher emission in the 304 \AA\ band since more condensations lie along the LOS. For better visibility, the maximum emission range of the right panel is increased from $2\,000 \, \mathrm{DN} \, \mathrm{s}^{-1}$ to $6\,000 \mathrm{DN} \, \mathrm{s}^{-1}$ in order to manifest the internal morphology of the central prominence. In the right panel,  which is at the moment when the mRTI fingers appear, it can be seen that the solar prominence clusters into the dip and creates an apparent monolithic structure at $z=0 \, \mathrm{Mm}$.  This is in agreement with observed solar prominences as seen from similar LOS viewpoints (see Fig.~1.i.  of \citet{berger2012}).  In our added movie,  blobs of coronal rain are spotted as well in the 304 \AA \ channel with both upwards- and downwards movements.  This is also in agreement with the results of \citet{li2022} who have simulated coronal rain in 2.5D and retrieved up- and downwards coronal rain blobs as well.  
 
  \begin{figure*}[!t]
 \begin{center}
    \includegraphics[width=\linewidth]{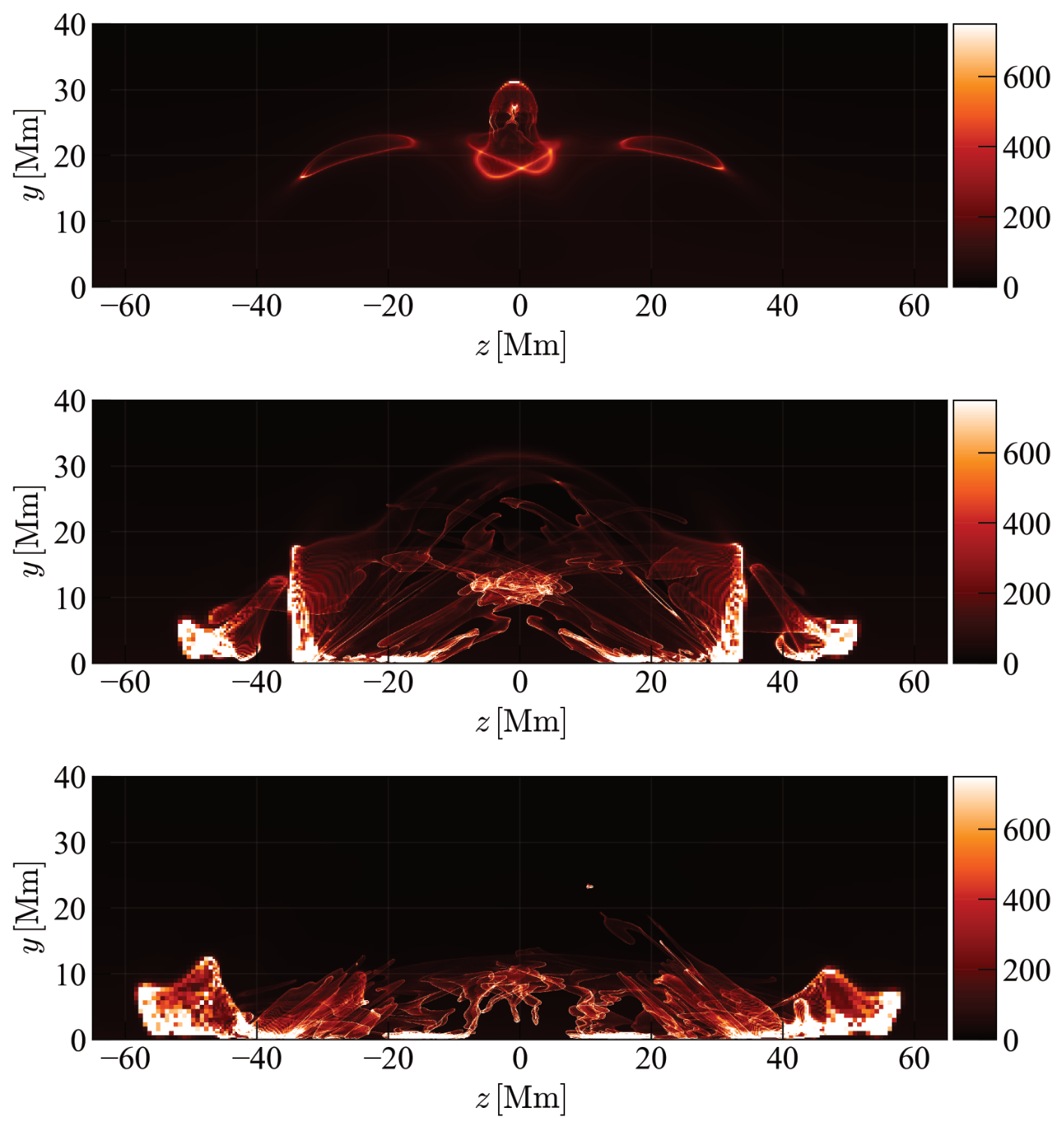}
 \end{center}
  \caption{Synthetic images of the solar prominence view as seen through the 304 \AA \ SDO/AIA EUV filter at the same times as in Fig.  \ref{fig:304_spine} but now the LOS is directed perpendicular to the flux rope. The top panel is captured at the onset of the prominence- and coronal rain formation, the middle panel during the falling towards the magnetic dips and the bottom panel when the central mRTI fingers are manifested and distinct.  Colours represent the datanumber per second.  The animation lasts for 21 seconds and shows how vertical structure is able to form in a horizontal magnetic field. (An animation of this figure is available in the online journal.)}\label{fig:304_prom}
 \end{figure*}
  
Other LOS angles provide complimentary insights on our synthetic solar prominence.  Fig.  \ref{fig:304_prom} shows the prominence at the same timestamps as in Fig.  \ref{fig:304_spine} but now the LOS is changed in order to have a side view of the entire structure.  From the top panel,  the formation of coronal rain occurs at a height of $y \approx 30 \, \mathrm{Mm}$ whereas the solar prominence forms at $y \approx 20 \, \mathrm{Mm}$.  The middle panel shows the snapshot when the prominence and coronal rain are more evolved. The entire structure fills out the domain $y \leqslant 35 \, \mathrm{Mm}$ with spontaneous in-situ condensations while the solar prominence is clustering at $y \approx 15 \, \mathrm{Mm}$ and $z \sim 0 \, \mathrm{Mm}$,  resulting in higher emission in 304 \AA.  In the added movie of the figure, coronal rain can be seen to glide along the magnetic field lines towards the bottom.  In the bottom panel,  the majority of the coronal rain has vanished through our open bottom boundary treatment, leaving only the solar prominence which extends to a height of $y \approx 14 \, \mathrm{Mm}$. The mRTI fingers are apparent and add vertical structure to an otherwise mostly horizontal magnetic field.  Vertical structures arising in solar prominences have been observed (see Fig.  1.h.  of \citet{su2012}) and recent numerical models have shown that the mRTI is able to bring in vertical structure \citep{jenkins2022}.  We confirm that mRTI arises naturally in solar prominences since denser condensations are suspended by the magnetic field above a lighter plasma which differs by two orders of magnitude in density. Only full 3D simulations can allow these mRTI perturbations to find the optimal minimal fieldline bending orientation to allow interchanges. We also note the presence of diverging horizontal flows away from the central prominence region  (which is more apparent in the movie). Thus plasma within the flux rope is still able to escape by gliding along magnetic field lines and we indeed obtain a draining,  quiescent solar prominence.

Quiescent prominences actually can exist in the solar corona for months, while we here demonstrate the formation and the MHD evolution of a quiescent prominence up to $9\,000$ seconds only, with the prominence fully formed for about $3\,500$ seconds. Again we emphasize that our simulation does not yet include lower photospheric and chromospheric dynamics, which the evaporation-condensation models invoke as an additional route for prominence formation and mass feeding. Within those $9\,000$ seconds, we noticed that some of the condensations are trapped within the magnetic flux rope and some glide away along strongly sheared, asymmetric, dipped magnetic arcades. Fig. \ref{fig:mass} shows that the simulated rate of mass drainage (bottom panel) steadily declines after $t\geqslant 7\,400 \, \mathrm{s}$. By a simple linear fit to this decline, we can estimate the order of magnitude of the lifespan. Note that the linear fit is performed for logarithmic $\dot{M}_\mathrm{drain}$ and time $t$ such that the decay of the prominence mass roughly follows an exponential trend.  We find that after $t\approx 34\,100 \, \mathrm{s}$ or about 9 hours the mass of the prominence will be depleted to zero.

   \begin{figure*}[!t]
 \begin{center}
  \includegraphics[width=\textwidth]{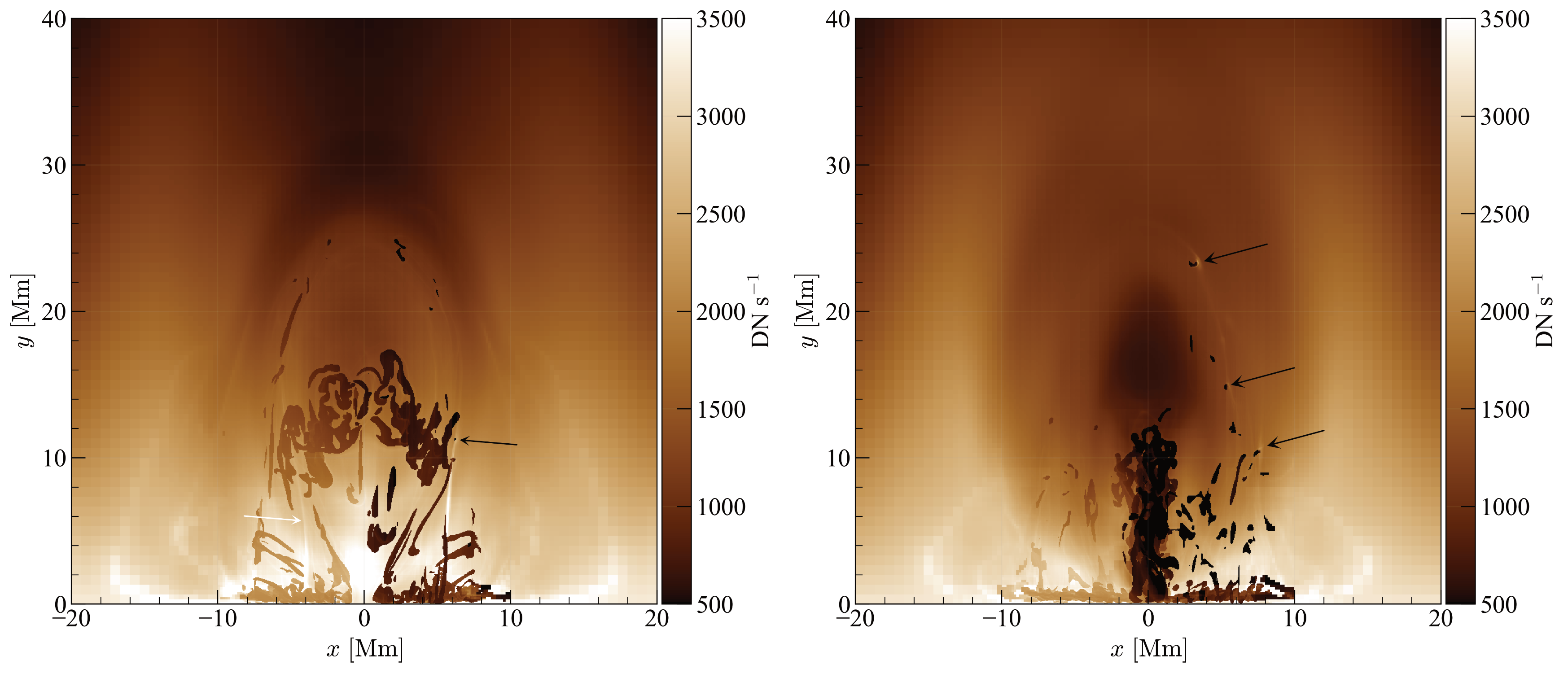}
 \end{center}
 \caption{Synthetic images of the spine of the solar prominence as seen through the 193 \AA \ SDO/AIA EUV filters. Condensations appear as dark structures. The left panel is captured at the moment when the condensations fall towards the magnetic dips.   The right panel exhibits the snapshot when the mRTI fingers collect in the dip.  The arrows point to the occurrence of reconnection outflows moving rapidly along magnetic field lines: black arrows indicate condensation-induced reconnection outflows and white arrows reconnection outflows without heavy condensations.    Colours represent the datanumber per second.  The animation lasts for 21 seconds and shows the formation of the flux rope and how coronal rain gives rise to reconnection outflow. (An animation of this figure is available in the online journal.)} \label{fig:193_spine}
 \end{figure*}

\subsubsection*{193 \AA\ EUV Channel}

Up to now, only the synthetic data of the 304 \AA\ emission band have been discussed. Fig.~\ref{fig:193_spine} shows the prominence at the same orientation and timestamps as Fig.~\ref{fig:304_spine} but now in the 193 \AA\ EUV wavelength band. Only hot plasma emits light in this wavelength channel,   cold condensations appear in absorption. To mimic optically thick effects of solar prominences in the 193 \AA\ EUV channel, we adopt the same procedure as \citet{xia2014} where emissions behind any condensation that obey a threshold inspired by optical depth estimates ($T \leqslant 0.015 \, \mathrm{MK}$ \& $n \geqslant 2.5 \cdot 10^{10} \, \mathrm{cm}^{-3}$) are excluded from the integration. In addition, due to our large domain the integration is only executed in the region of interest where the prominence resides, i.e. $\vert y \vert < 40 \, \mathrm{Mm}$ and $\vert z \vert < 65 \, \mathrm{Mm}$. The left panel corresponds to a moment when overarching coronal rain as well as flux-rope embedded prominence condensations start to collect towards dips, and it also reveals a transient phenomenon. Narrow and bursty emission sites appear as indicated by the white and black arrows, depending whether the outflow is induced by a condensation (black arrow) or not (white arrow). As we will show shortly, these localized brightenings signal reconnection outflows from local magnetic reconnection events at specific locations within the simulation domain. The right panel representative of the fully formed prominence structure only, shows the presence of a coronal cavity which is typically associated with solar prominences \citep{gibson2014}.  \citet{gibson2010} have shown that the density within coronal cavities is lower than the surrounding coronal environment. We indeed find that after the solar prominence has formed, the density within the flux rope drops from $1.5 \cdot 10^{-15} \,  \mathrm{g} \, \mathrm{cm}^{-3}$ to $6.2 \cdot 10^{-16} \,  \mathrm{g} \, \mathrm{cm}^{-3}$.

\begin{figure*}[!t]
 \begin{center}
    \includegraphics[width=\linewidth]{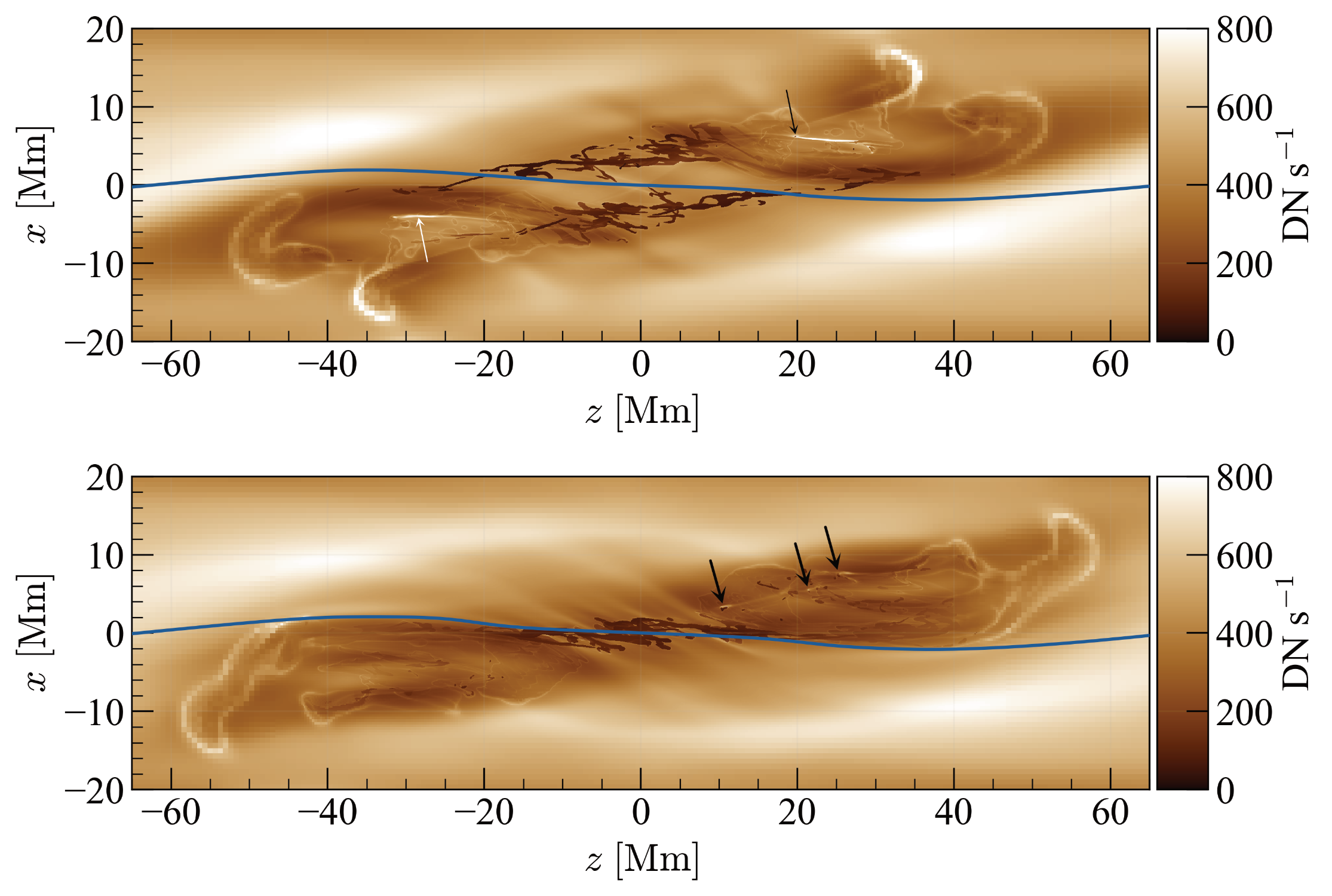}
 \end{center}
 \caption{Synthetic images looking down on the spine of the solar prominence as seen through the 193 \AA \ SDO/AIA EUV filters from the top. The blue curve tracks the location of the PIL. Both panels are captured when localized, transient reconnection outflows are detected: black arrows indicate condensation-induced reconnection outflows and white arrows reconnection outflows without heavy condensations. The bottom panel is shown when the mRTI fingers cluster into a monolithic structure from the plane-of-sky angle. Colours represent the datanumber per second.  The animation lasts for 13 seconds and starts after condensations have occurred ($T \leqslant 0.1 \, \mathrm{MK}$). It shows the difference in movement between the coronal rain and prominence. (An animation of this figure is available in the online journal.)} \label{fig:193_fil}
 \end{figure*}

Fig.~\ref{fig:193_fil} displays the same snapshots as Fig.~\ref{fig:193_spine} but now along a different LOS that views the prominence structure from the top, i.e. the filament view. The polarity inversion line (PIL) is shown by the blue curve. Condensations are seen as dark regions where absorption is high. Bright regions relate to hot plasma. The bottom panel shows the solar prominence lying parallel to the PIL which meets the observational constraint that solar prominences demonstrate only a small angle with respect to the polarity inversion line \citep{martin1998}. In the top panel, two narrow and elongated high temperature regions emit strongly and they are indicated by the white arrows. These are related to spontaneous magnetic reconnection events. The temperature and velocity of the upper reconnection outflow is measured to be about  $T \approx 2.7 \,  \mathrm{MK}$ and $v \approx 252 \,  \mathrm{km} \, \mathrm{s}^{-1}$,  respectively. Four reconnection outflows are detected in total, two of which are induced by coronal rain and are seen in both panels of Fig. ~\ref{fig:193_spine} and ~\ref{fig:193_fil} pointed by the black arrows. 

\begin{figure*}[!t]
 \begin{center}
\includegraphics[width=\linewidth]{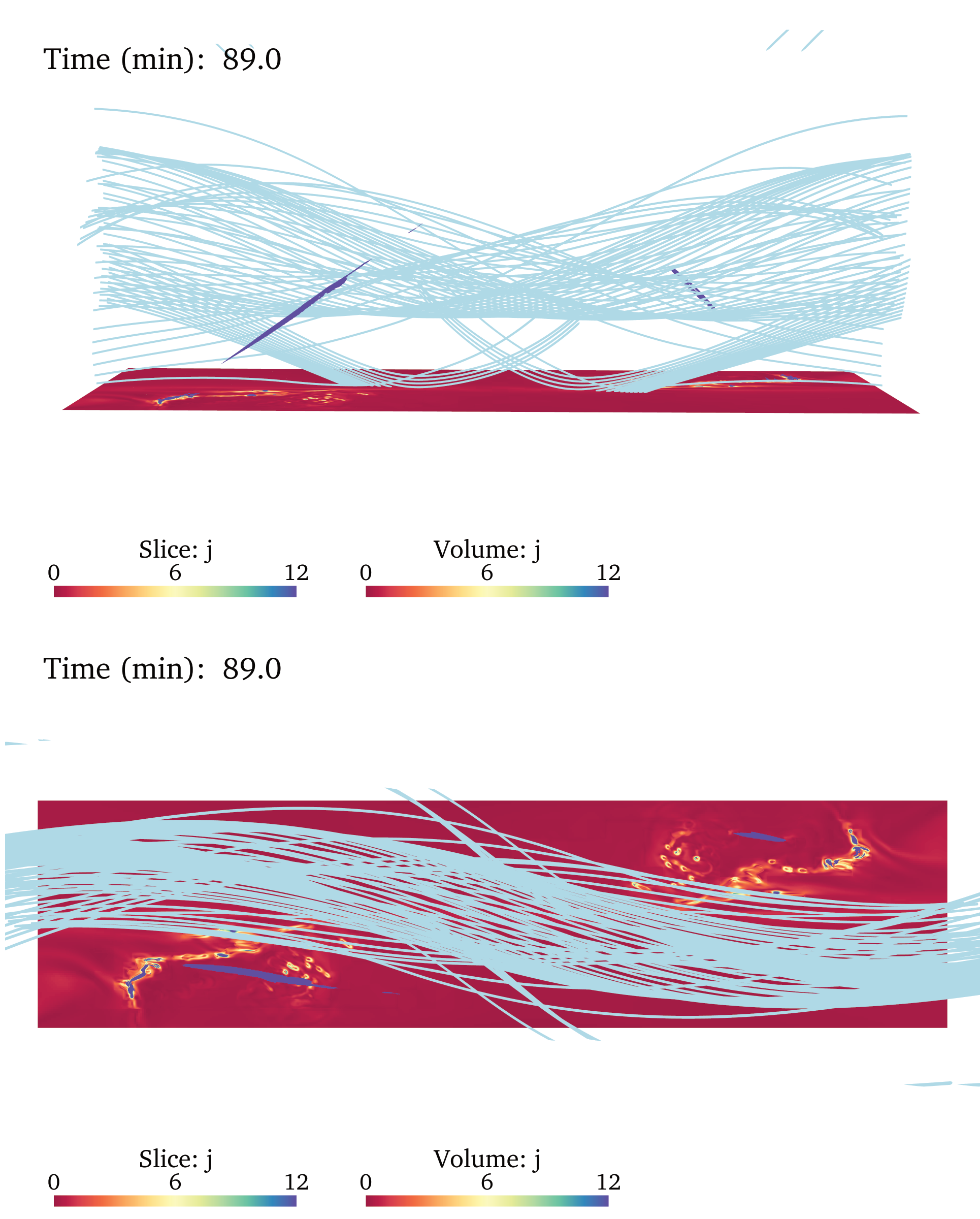}
 \end{center}
 \caption{Magnitude of the current density $J$ in code units,  i.e.  $1.59 \cdot 10^{-9} \,  \mathrm{statA} \, \mathrm{cm}^{-2}$ as seen from the top (top panel) and side (bottom panel) with superimposed magnetic field lines. Both panels are at the same timestamp as the top panel of Fig.~\ref{fig:193_fil}. The bottom plane shows the magnitude of the current density and the volume rendering (dark blue) identifies those locations within the grid where the local current density magnitude exceeds the threshold $J \geqslant 12$. Only the domain $\vert x \vert \leqslant 10 \, \mathrm{Mm}, y \leqslant 35 \, \mathrm{Mm}$ and $\vert z \vert \leqslant 65 \, \mathrm{Mm}$ is shown.}\label{fig:top_current}
 \end{figure*}

To show that the narrow and transient bright flashes in the 193 \AA\ channel are indeed magnetic reconnection events, Fig.~\ref{fig:top_current} shows the magnitude of the current density $J$ in code units $1.59 \cdot 10^{-9} \,  \mathrm{statA} \, \mathrm{cm}^{-2}$ at the location where the reconnection outflows originate from. Representative magnetic field lines are superimposed as well. The volume rendering identifies those regions where $J > 12$ in code units. There, strong current sheets are created that reach large values up to $J \approx 50$ or $8.0 \cdot 10^{-8} \,  \mathrm{statA} \, \mathrm{cm}^{-2}$, resulting in spontaneous magnetic reconnection events. Note that these specific magnetic reconnections arise due to numerical resistivity as our user-specified resistivity is turned off at this stage. These magnetic reconnection events do not affect the solar prominence since the reconnected magnetic field lines from these current sheets do not intercept with the prominence. The current variations shown on the bottom plane correspond to those locations where condensations touch the bottom boundary, as these are also accompanied by current concentrations (as shown for the mRTI fingers in Fig.~\ref{fig:current_densities}). 

These findings on bursty reconnections and associated brightenings are in almost perfect agreement with \citet{reeves2015} who have observed reconnection outflows around an erupting prominence which were prominently seen through the 171 \AA\ SDO/AIA EUV channel and the \textit{Interface Region Imaging Spectrometer} (IRIS) satellite. They concluded that these outflows are indeed caused by a local magnetic reconnection event occurring between the prominence magnetic field pushing against the overlying arcades as it erupts. The observed reconnection outflows reached a temperature of $2.5 \, \mathrm{MK}$ and velocities from the IRIS data along the LOS varied between $112 -197 \, \mathrm{km} \, \mathrm{s}^{-1}$ whereas from the AIA data in the plane-of-sky the velocity was observed to be about $300 \, \mathrm{km} \, \mathrm{s}^{-1}$. In our case two reconnection outflows manifestly arise from heavy condensations. We have shown earlier that condensations indeed inherit strong gradients in current density (see Fig. ~\ref{fig:current_densities}) which explains why their local magnetic fields would burstly reconnect. The origin of the two other reconnection outflows are not manifested as they are blocked by overlying coronal rain, making us unable to conclude whether they are induced by condensations as well.

\section*{Summary and outlook}
Our high-resolution, 3D MHD simulated draining quiescent prominence, resolving prominence details at 41.7 km (with a zoom-in local run to 20 km), has tackled various open questions in the prominence research field. We especially clarified the dynamic mass cycle of a quiescent prominence when there is no role for evaporation processes from a dense underlying chromosphere.  While we have established the presence of siphoning flows, no correlation was found between the prominence mass within a magnetic field thread $\mathcal{M}_\mathrm{line}$ and the footpoint flow velocity $\mathcal{V}_\mathrm{fp}$ due to the fact that the mass siphoning did not have enough time to traverse the magnetic field line within the simulation time. Nevertheless, we obtain mass siphoning without any active added heating prescription and in a purely coronal volume proves that thermal instability is sufficient to set up pressure variations that can overcome the force of gravity and induce these flows. This also implies that a localised heating source at the footpoints of the magnetic field is not a necessary condition to promote siphoning flows. We self-consistently obtain tangent flows along the magnetic field line in the range $0 - 35 \,  \mathrm{km} \, \mathrm{s}^{-1}$ which arise from the pressure gradient.  For the first time, these siphoning flows have been shown explicitly by conducting a statistical study of $1 \,209$ magnetic field lines which contain cold condensations (see Fig. \ref{fig:pearson}). Even without a chromosphere-transition region included but with clear PCTR variations in the coronal volume arising naturally,  we obtain total prominence mass values that lie in the range of observations. Future work that extends our simulation down to lower heights and increase the simulation time to allow the siphoning flows to reach the prominences will only raise the masses obtained, consistent with our model representing the lower observational mass range.

We also retrieved coronal rain forming above the flux rope and gliding down along arcade field lines, to disappear out of our simulation domain.  Other numerical work that simulated coronal rain relied on ad-hoc localised heating rate prescriptions in order to form them.  We have shown that in the reconnection-condensation model coronal rain forms due to the high ram pressure exerted by the rising flux rope during its formation.  This is the first time that a solar prominence and coronal rain are obtained simultaneously in a 3D simulation. A 2.5D variant of a combined solar prominence and coronal rain has been recently obtained as well \citep{veronika2024}. 

In our simulation, we identified clear distinctive roles for the three magnetic topologies that are often separated in prominence literature, i.e. flux rope magnetic field lines, non-dipped arcades and dipped, highly-sheared and asymmetric dipped arcades. Each class of these topologies carry a distinct role regarding condensation morphology and evolution: the non-dipped arcades are associated with coronal rain, the flux rope suspends the prominence in the solar corona and sheared and dipped magnetic arcades allow for prominence material to escape the flux rope. We provided a full picture on the mass cycle of quiescent solar prominences in our self-consistent simulation: the origin of its mass and how the mass drains. \citet{xia2014} adopted the evaporation-condensation model at a lower resolution and obtained a dynamic solar prominenence with mass drainage as well. Their drainage was a result of local condensations dragging the magnetic field lines downwards and condensations that formed at the legs of arched magnetic arcades falling down. While our works use essentially similar physical and numerical ingredients, there are two key differences. The first one is that their mass siphoning originates from localised footpoint heating whereas ours arises purely from thermal instability. The second is that our draining condensations do not form at the legs of the arched arcades. Instead, our draining condensations occur close to the apex of sheared and asymmetric dipped magnetic arcades, allowing them to convert their large gravitational energy to kinetic energy. These condensations gain speed as they fall down to overshoot the dip and continue to glide to the lower corona. Because our estimates of the condensation growth rate $\langle \dot{M}_\mathrm{cond} \rangle$ and mass drainage rate $\langle \dot{M}_\mathrm{drain}\rangle$ agree very well with observations of a draining quiescent solar prominence made by \citet{liu2012}, we conclude that our self-consistent model is able to capture all the important aspects of the mass cycle of a quiescent solar prominence.

The prominence further undergoes the magnetic Rayleigh-Taylor instability which gives rise to a vertical structure in an otherwise horizontal magnetic field. By quantifying the vertical forces, the vertical velocity and the density of two fingers,  we have shown that magnetic tension is the primary decelerator of the fingers and magnetic tension and gravity the main accelerators.  By increasing the effective resolution from 41.7 km to 20.8 km, we found a tendency to even finer Rayleigh-Taylor details.

Lastly, we have synthesised our data which shows remarkable agreement with observed solar prominences.  We obtain the coronal cavity that is typically associated with solar prominences, 
 vertical structure from the magnetic Rayleigh-Taylor instability and a prominence that lies parallel with the polarity inversion line. From the 193 \AA\ EUV channel,  reconnection outflows have been self-consistently obtained for the first time in solar prominence simulations that agree very well with observations made by \citet{reeves2015}. Two of the four reconnection outflows are induced by coronal rain. Because our results are able to draw many parallels to observations, we conclude that the current reconnection-condensation model truly resembles flux rope and prominence formation processes.
 
Our results show that, despite various idealizations, like the purely optically thin radiation assumption,  we are able to relate to observations to a very satisfactory degree. We infer that the addition of a chromosphere which -- when heated locally -- may serve as an extra mass supply into the corona will result in higher mass values that can then also explain the origin of large prominences with masses in the order of $10^{14} - 10^{15} \,  \mathrm{g}$. \citet{zhou2020} showed that if this localised heating is randomized, counterstreaming flows arise naturally which have been observed in solar filaments as well \citep{diercke2018}. The lifespan of our solar prominence is in the order of hours, while actual prominences can live weeks to months. On those longer timescales, the coupling and extra mass cycling with the lower atmospheric layers will help sustain long-lived prominences. In our idealized numerical simulations, we still find reconnection and overall mass slippage which are treated by diffusive processes. By using more higher-order flux limiters, mass slippage is better constrained and consequently, the prominence will be able to suspend for longer periods. In addition to these enhancements, \citet{liu2022} incorporated convective motions of supergranules in their simulation bottom boundary and found that a magnetic flux rope is able to form if the Coriolis force is included in the MHD equations. Since our flux rope forms under much more ideal conditions, the next step would be to simulate the bottom boundary more realistically by including these effects as well. Incorporating these improvements will result in further agreement with observations. Future work can concentrate on synthetic spectroscopic views on the 3D prominence, which has been pioneered recently by \citet{jenkins2023}.

\subsection*{Acknowledgements}
DD  would like to thank Jack Jenkins, Joris Hermans, Fabio Bacchini, Tine Baelmans,  Valeriia Liakh and Ehsan Moravveji for the insightful discussions on improvements of the code or results. The tools needed for the visualisations have been facilitated by using Python (www.python.org) and the inner modules yt-project \citep{yt},  PyVista \citep{pyvista} and matplotlib \citep{matplotlib}.
RK is supported by Internal Funds KU Leuven through the project C14/19/089 TRACESpace and an FWO project G0B4521N. 
We acknowledge funding from the European Research Council (ERC) under the European Union Horizon 2020 research and innovation program (grant agreement No. 833251 PROMINENT ERC-ADG 2018). The computational resources and services used in this work were provided by the VSC (Flemish Supercomputer Center), funded by the Research Foundation Flanders (FWO) and the Flemish Government, department EWI.


\newpage
\appendix

\section{Estimation of the Mass Reservoir within the Flux Rope}

In Section 3 we have shown that the total mass of the solar prominence exceeds the mass reservoir within the formed flux rope. Here we elaborate on how this mass reservoir has been estimated. 

 \begin{figure*}[!h]
 \begin{center}
  \includegraphics[width=\textwidth]{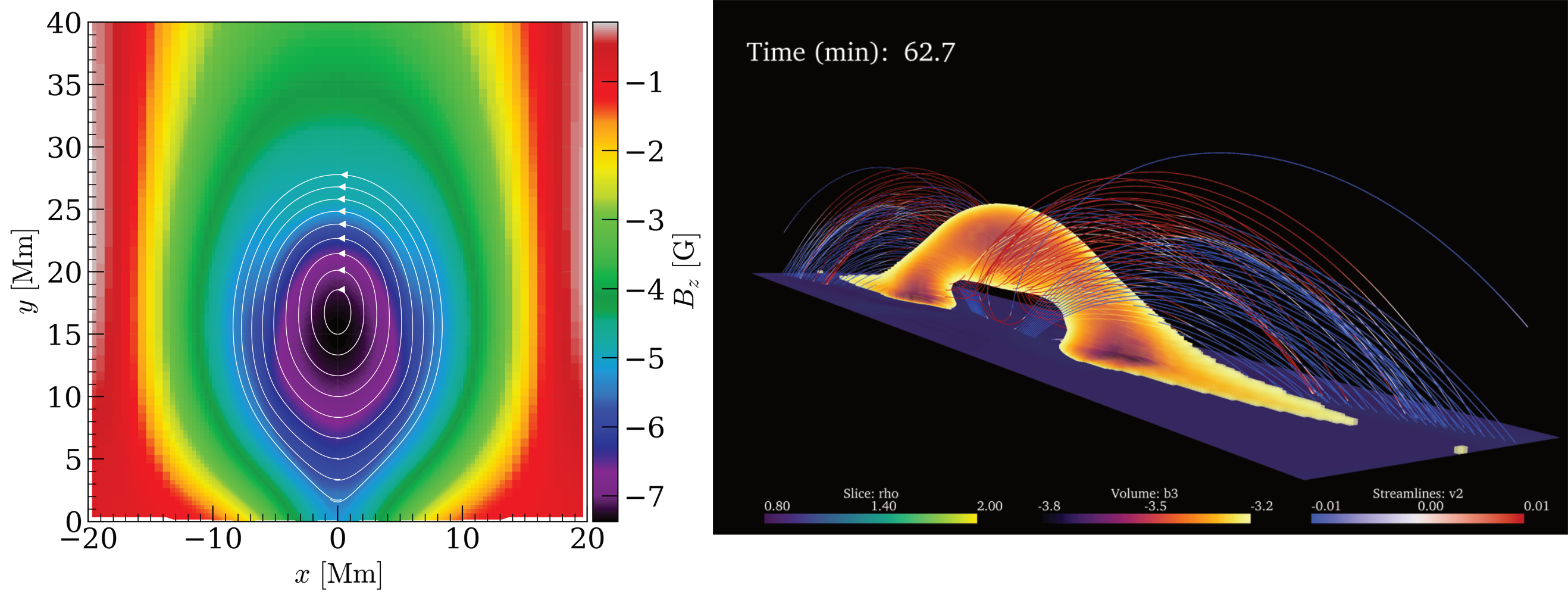}
 \caption{Left panel displays the distribution of $B_z$ within the $z=0$ plane with overlaid poloidal magnetic field lines.   The right panel shows the flux rope's interior by defining regions in space that fulfil the condition $B_z \leqslant -6.4 \, \mathrm{G}$ or -3.2 in code units. } \label{fig:fluxrope_interior}
 \end{center}
 \end{figure*}
 
Fig.~\ref{fig:fluxrope_interior} shows the $B_z$ distribution in the central vertical $z=0$ plane in the left panel and the three dimensional volume for which $B_z \leqslant -6.4 \,  \mathrm{G}$ at right. The magnetic flux rope is here well approximated by different $B_z$-isosurfaces and we have found that the isosurface $B_z = -6.4 \,  \mathrm{G}$ encompasses the interior of the flux rope to a satisfactory degree at the moment before condensations occur. Alternatively, the same isosurface can be retrieved by iteratively varying over all vertical $z$-planes and searching for the outermost, closed poloidal fieldline for each plane to create a closed flux rope surface. Since the condition $B_z \leqslant -6.4 \,  \mathrm{G}$ coincides nicely with the flux rope, only those cells in the simulation domain are selected that satisfy this constraint.  Once the relevant cells have been found, the mass is simply calculated as 
\begin{equation}
M_\mathrm{fr} = \iiint_{\mathcal{V}_\mathrm{fr}} \rho \mathrm{d}V
\end{equation}
 with ${\mathcal{V}_\mathrm{fr}}$ the volume of the flux rope that satisfies $B_z \leqslant -6.4 \,  \mathrm{G}$.  The mass value at $t= 3 \, 761 \,  \mathrm{s}$,  i.e.  the moment before condensations occur ($T \leqslant 0.1 \,  \mathrm{MK}$),  is chosen as the mass reservoir of the flux rope since the flux rope is then unaffected by coronal rain and has evolved sufficiently enough for its growth to stagnate.  This results in our estimated mass reservoir of the flux rope equal to $M_\mathrm{fr} = 1.9 \cdot 10^{13} \,  \mathrm{g}$.


\bibliography{sample631}{}

\begin{thebibliography}{}
\expandafter\ifx\csname natexlab\endcsname\relax\def\natexlab#1{#1}\fi
\providecommand{\url}[1]{\href{#1}{#1}}
\providecommand{\dodoi}[1]{doi:~\href{http://doi.org/#1}{\nolinkurl{#1}}}
\providecommand{\doeprint}[1]{\href{http://ascl.net/#1}{\nolinkurl{http://ascl.net/#1}}}
\providecommand{\doarXiv}[1]{\href{https://arxiv.org/abs/#1}{\nolinkurl{https://arxiv.org/abs/#1}}}

\bibitem[{{Antiochos} \& {Klimchuk}(1991)}]{antiochos1991}
{Antiochos}, S.~K., \& {Klimchuk}, J.~A. 1991, \apj, 378, 372, \dodoi{10.1086/170437}

\bibitem[{{Antolin} \& {Froment}(2022)}]{antolin2022b}
{Antolin}, P., \& {Froment}, C. 2022, Frontiers in Astronomy and Space Sciences, 9, 820116, \dodoi{10.3389/fspas.2022.820116}

\bibitem[{Antolin {et~al.}(2022)Antolin, Mart{\'\i}nez-Sykora, \& {\c{S}}ahin}]{antolin2022}
Antolin, P., Mart{\'\i}nez-Sykora, J., \& {\c{S}}ahin, S. 2022, The Astrophysical Journal Letters, 926, L29

\bibitem[{Antolin {et~al.}(2010)Antolin, Shibata, \& Vissers}]{antolin2010}
Antolin, P., Shibata, K., \& Vissers, G. 2010, The Astrophysical Journal, 716, 154

\bibitem[{Antolin \& van Der~Voort(2012)}]{antolin2012}
Antolin, P., \& van Der~Voort, L.~R. 2012, The Astrophysical Journal, 745, 152

\bibitem[{Berger {et~al.}(2012)Berger, Liu, \& Low}]{berger2012}
Berger, T.~E., Liu, W., \& Low, B. 2012, The Astrophysical Journal Letters, 758, L37

\bibitem[{Berger {et~al.}(2008)Berger, Shine, Slater, Tarbell, Okamoto, Ichimoto, Katsukawa, Suematsu, Tsuneta, Lites, {et~al.}}]{berger2008}
Berger, T.~E., Shine, R.~A., Slater, G.~L., {et~al.} 2008, The Astrophysical Journal, 676, L89

\bibitem[{Berger {et~al.}(2010)Berger, Slater, Hurlburt, Shine, Tarbell, Lites, Okamoto, Ichimoto, Katsukawa, Magara, {et~al.}}]{berger2010}
Berger, T.~E., Slater, G., Hurlburt, N., {et~al.} 2010, The Astrophysical Journal, 716, 1288

\bibitem[{Braileanu \& Keppens(2021)}]{braileanu2021}
Braileanu, B.~P., \& Keppens, R. 2021, Astronomy \& Astrophysics, 653, A131

\bibitem[{Brughmans {et~al.}(2022)Brughmans, Jenkins, \& Keppens}]{brughmans2022}
Brughmans, N., Jenkins, J., \& Keppens, R. 2022, Astronomy \& Astrophysics, 668, A47

\bibitem[{Claes {et~al.}(2020)Claes, Keppens, \& Xia}]{claes2020}
Claes, N., Keppens, R., \& Xia, C. 2020, Astronomy \& Astrophysics, 636, A112

\bibitem[{{Diercke} {et~al.}(2018){Diercke}, {Kuckein}, {Verma}, \& {Denker}}]{diercke2018}
{Diercke}, A., {Kuckein}, C., {Verma}, M., \& {Denker}, C. 2018, \aap, 611, A64, \dodoi{10.1051/0004-6361/201730536}

\bibitem[{Fang {et~al.}(2013)Fang, Xia, \& Keppens}]{fang2013}
Fang, X., Xia, C., \& Keppens, R. 2013, The Astrophysical Journal Letters, 771, L29

\bibitem[{Fang {et~al.}(2015)Fang, Xia, Keppens, \& Van~Doorsselaere}]{fang2015}
Fang, X., Xia, C., Keppens, R., \& Van~Doorsselaere, T. 2015, The Astrophysical Journal, 807, 142

\bibitem[{Field(1965)}]{field1965}
Field, G.~B. 1965, Astrophysical Journal, vol. 142, p. 531, 142, 531

\bibitem[{{Froment} {et~al.}(2018){Froment}, {Auch{\`e}re}, {Miki{\'c}}, {Aulanier}, {Bocchialini}, {Buchlin}, {Solomon}, \& {Soubri{\'e}}}]{froment2018}
{Froment}, C., {Auch{\`e}re}, F., {Miki{\'c}}, Z., {et~al.} 2018, \apj, 855, 52, \dodoi{10.3847/1538-4357/aaaf1d}

\bibitem[{Gibson(2014)}]{gibson2014}
Gibson, S. 2014, in Solar prominences (Springer), 323--353

\bibitem[{Gibson {et~al.}(2010)Gibson, Kucera, Rastawicki, Dove, De~Toma, Hao, Hill, Hudson, Marqu{\'e}, McIntosh, {et~al.}}]{gibson2010}
Gibson, S.~E., Kucera, T., Rastawicki, D., {et~al.} 2010, The Astrophysical Journal, 724, 1133

\bibitem[{Goedbloed {et~al.}(2019)Goedbloed, Goedbloed, Keppens, \& Poedts}]{goedbloed2019}
Goedbloed, H., Goedbloed, J., Keppens, R., \& Poedts, S. 2019, Magnetohydrodynamics: Of Laboratory and Astrophysical Plasmas (Cambridge University Press)

\bibitem[{Gudiksen {et~al.}(2011)Gudiksen, Carlsson, Hansteen, Hayek, Leenaarts, \& Mart{\'\i}nez-Sykora}]{bifrost}
Gudiksen, B.~V., Carlsson, M., Hansteen, V.~H., {et~al.} 2011, Astronomy \& Astrophysics, 531, A154

\bibitem[{Guo {et~al.}(2010)Guo, Schmieder, D{\'e}moulin, Wiegelmann, Aulanier, T{\"o}r{\"o}k, \& Bommier}]{guo2010}
Guo, Y., Schmieder, B., D{\'e}moulin, P., {et~al.} 2010, The Astrophysical Journal, 714, 343

\bibitem[{Hermans \& Keppens(2021)}]{hermans2021}
Hermans, J., \& Keppens, R. 2021, Astronomy \& Astrophysics, 655, A36

\bibitem[{Hildner(1974)}]{hildner1974}
Hildner, E. 1974, Solar Physics, 35, 123

\bibitem[{Hunter(2007)}]{matplotlib}
Hunter, J.~D. 2007, Computing in science \& engineering, 9, 90

\bibitem[{Jenkins \& Keppens(2021)}]{jenkins2021}
Jenkins, J., \& Keppens, R. 2021, Astronomy \& Astrophysics, 646, A134

\bibitem[{Jenkins \& Keppens(2022)}]{jenkins2022}
---. 2022, Nature Astronomy, 6, 942

\bibitem[{{Jenkins} {et~al.}(2023){Jenkins}, {Osborne}, \& {Keppens}}]{jenkins2023}
{Jenkins}, J.~M., {Osborne}, C.~M.~J., \& {Keppens}, R. 2023, \aap, 670, A179, \dodoi{10.1051/0004-6361/202244868}

\bibitem[{Jer{\v{c}}i{\'c} \& Keppens(2022)}]{jervcic2022}
Jer{\v{c}}i{\'c}, V., \& Keppens, R. 2022, arXiv preprint arXiv:2212.08537

\bibitem[{{Jer\v{c}i\'{c}} {et~al.}(Submitted){Jer\v{c}i\'{c}}, M., \& {Keppens}}]{veronika2024}
{Jer\v{c}i\'{c}}, V., M., J.~J., \& {Keppens}, R. Submitted, Submitted to A\&A

\bibitem[{{Johnston} {et~al.}(2019){Johnston}, {Cargill}, {Antolin}, {Hood}, {De Moortel}, \& {Bradshaw}}]{johnston2019}
{Johnston}, C.~D., {Cargill}, P.~J., {Antolin}, P., {et~al.} 2019, \aap, 625, A149, \dodoi{10.1051/0004-6361/201834742}

\bibitem[{Kaneko \& Yokoyama(2017)}]{kaneko2017}
Kaneko, T., \& Yokoyama, T. 2017, The Astrophysical Journal, 845, 12

\bibitem[{Kaneko \& Yokoyama(2018)}]{kaneko2018}
---. 2018, The Astrophysical Journal, 869, 136

\bibitem[{Keppens {et~al.}(2023)Keppens, Braileanu, Zhou, Ruan, Xia, Guo, Claes, \& Bacchini}]{amrvac3.0}
Keppens, R., Braileanu, B.~P., Zhou, Y., {et~al.} 2023, arXiv preprint arXiv:2303.03026

\bibitem[{Keppens \& Xia(2014)}]{keppens2014}
Keppens, R., \& Xia, C. 2014, The Astrophysical Journal, 789, 22

\bibitem[{{Klimchuk} \& {Luna}(2019)}]{klimchuck2019}
{Klimchuk}, J.~A., \& {Luna}, M. 2019, \apj, 884, 68, \dodoi{10.3847/1538-4357/ab41f4}

\bibitem[{Kohutova {et~al.}(2020)Kohutova, Antolin, Popovas, Szydlarski, \& Hansteen}]{kohutova2020}
Kohutova, P., Antolin, P., Popovas, A., Szydlarski, M., \& Hansteen, V. 2020, Astronomy \& Astrophysics, 639, A20

\bibitem[{Labrosse {et~al.}(2010)Labrosse, Heinzel, Vial, Kucera, Parenti, Gun{\ss}r, Schmieder, \& Kilper}]{labrosse2010}
Labrosse, N., Heinzel, P., Vial, J.-C., {et~al.} 2010, Space Science Reviews, 151, 243

\bibitem[{Lemen {et~al.}(2012)Lemen, Title, Akin, Boerner, Chou, Drake, Duncan, Edwards, Friedlaender, Heyman, {et~al.}}]{sdoaia}
Lemen, J.~R., Title, A.~M., Akin, D.~J., {et~al.} 2012, Solar Physics, 275, 17

\bibitem[{Li {et~al.}(2022)Li, Keppens, \& Zhou}]{li2022}
Li, X., Keppens, R., \& Zhou, Y. 2022, The Astrophysical Journal, 926, 216

\bibitem[{Lin(2011)}]{lin2011}
Lin, Y. 2011, Space science reviews, 158, 237

\bibitem[{{Liu} \& {Xia}(2022)}]{liu2022}
{Liu}, Q., \& {Xia}, C. 2022, \apjl, 934, L9, \dodoi{10.3847/2041-8213/ac80c6}

\bibitem[{{Liu} {et~al.}(2012){Liu}, {Berger}, \& {Low}}]{liu2012}
{Liu}, W., {Berger}, T.~E., \& {Low}, B.~C. 2012, \apjl, 745, L21, \dodoi{10.1088/2041-8205/745/2/L21}

\bibitem[{Low {et~al.}(2012)Low, Liu, Berger, \& Casini}]{low2012}
Low, B., Liu, W., Berger, T., \& Casini, R. 2012, The Astrophysical Journal, 757, 21

\bibitem[{Luna {et~al.}(2012)Luna, Karpen, \& DeVore}]{luna2012}
Luna, M., Karpen, J., \& DeVore, C. 2012, The Astrophysical Journal, 746, 30

\bibitem[{Mackay(2021)}]{mackay2021}
Mackay, D.~H. 2021, Oxford Research Encyclopedia of Physics

\bibitem[{Martin(1998)}]{martin1998}
Martin, S.~F. 1998, Solar Physics, 182, 107

\bibitem[{Moschou {et~al.}(2015)Moschou, Keppens, Xia, \& Fang}]{moschou2015}
Moschou, S., Keppens, R., Xia, C., \& Fang, X. 2015, Advances in Space Research, 56, 2738

\bibitem[{M{\"u}ller {et~al.}(2003)M{\"u}ller, Hansteen, \& Peter}]{muller2003}
M{\"u}ller, D., Hansteen, V., \& Peter, H. 2003, Astronomy \& Astrophysics, 411, 605

\bibitem[{{M{\"u}ller} {et~al.}(2005){M{\"u}ller}, {De Groof}, {Hansteen}, \& {Peter}}]{muller2005}
{M{\"u}ller}, D.~A.~N., {De Groof}, A., {Hansteen}, V.~H., \& {Peter}, H. 2005, \aap, 436, 1067, \dodoi{10.1051/0004-6361:20042141}

\bibitem[{{Parker}(1953)}]{parker1953}
{Parker}, E.~N. 1953, \apj, 117, 431, \dodoi{10.1086/145707}

\bibitem[{Pelouze {et~al.}(2022)Pelouze, Auch{\`e}re, Bocchialini, Froment, Miki{\'c}, Soubri{\'e}, \& Voyeux}]{pelouze2022}
Pelouze, G., Auch{\`e}re, F., Bocchialini, K., {et~al.} 2022, Astronomy \& Astrophysics, 658, A71

\bibitem[{Poland \& Mariska(1986)}]{poland1986}
Poland, A., \& Mariska, J.~T. 1986, Solar physics, 104, 303

\bibitem[{{Reeves} {et~al.}(2015){Reeves}, {McCauley}, \& {Tian}}]{reeves2015}
{Reeves}, K.~K., {McCauley}, P.~I., \& {Tian}, H. 2015, \apj, 807, 7, \dodoi{10.1088/0004-637X/807/1/7}

\bibitem[{{\c{S}}ahin {et~al.}(2023){\c{S}}ahin, Antolin, Froment, \& Schad}]{sahin2023}
{\c{S}}ahin, S., Antolin, P., Froment, C., \& Schad, T.~A. 2023, The Astrophysical Journal, 950, 171

\bibitem[{{Secchi}(1875)}]{secchi1875}
{Secchi}, A. 1875, {Le Soleil}, \dodoi{10.3931/e-rara-14748}

\bibitem[{Smith \& Priest(1977)}]{smith1977}
Smith, E., \& Priest, E. 1977, Solar Physics, 53, 25

\bibitem[{Spitzer(2006)}]{spitzer2006}
Spitzer, L. 2006, Physics of fully ionized gases (Courier Corporation)

\bibitem[{Su \& Van~Ballegooijen(2012)}]{su2012}
Su, Y., \& Van~Ballegooijen, A. 2012, The Astrophysical Journal, 757, 168

\bibitem[{Sullivan \& Kaszynski(2019)}]{pyvista}
Sullivan, C., \& Kaszynski, A. 2019, Journal of Open Source Software, 4, 1450

\bibitem[{Terradas {et~al.}(2015)Terradas, Soler, Luna, Oliver, \& Ballester}]{terradas2015}
Terradas, J., Soler, R., Luna, M., Oliver, R., \& Ballester, J. 2015, The Astrophysical Journal, 799, 94

\bibitem[{{Turk} {et~al.}(2011){Turk}, {Smith}, {Oishi}, {Skory}, {Skillman}, {Abel}, \& {Norman}}]{yt}
{Turk}, M.~J., {Smith}, B.~D., {Oishi}, J.~S., {et~al.} 2011, The Astrophysical Journal Supplement Series, 192, 9, \dodoi{10.1088/0067-0049/192/1/9}

\bibitem[{Xia {et~al.}(2012)Xia, Chen, \& Keppens}]{xia2012}
Xia, C., Chen, P., \& Keppens, R. 2012, The Astrophysical Journal Letters, 748, L26

\bibitem[{Xia {et~al.}(2011)Xia, Chen, Keppens, \& van Marle}]{xia2011}
Xia, C., Chen, P., Keppens, R., \& van Marle, A.~J. 2011, The Astrophysical Journal, 737, 27

\bibitem[{Xia \& Keppens(2016)}]{xia2016}
Xia, C., \& Keppens, R. 2016, The Astrophysical Journal Letters, 825, L29

\bibitem[{Xia {et~al.}(2014)Xia, Keppens, Antolin, \& Porth}]{xia2014}
Xia, C., Keppens, R., Antolin, P., \& Porth, O. 2014, The Astrophysical Journal Letters, 792, L38

\bibitem[{Xia {et~al.}(2017)Xia, Keppens, \& Fang}]{xia2017}
Xia, C., Keppens, R., \& Fang, X. 2017, Astronomy \& Astrophysics, 603, A42

\bibitem[{Yang {et~al.}(2021)Yang, Yang, Bi, Hong, \& Xu}]{yang2021}
Yang, B., Yang, J., Bi, Y., Hong, J., \& Xu, Z. 2021, The Astrophysical Journal Letters, 921, L33

\bibitem[{Zhou {et~al.}(2020)Zhou, Chen, Hong, \& Fang}]{zhou2020}
Zhou, Y., Chen, P., Hong, J., \& Fang, C. 2020, Nature Astronomy, 4, 994

\bibitem[{{Zhou} {et~al.}(2014){Zhou}, {Chen}, {Zhang}, \& {Fang}}]{zhou2014}
{Zhou}, Y., {Chen}, P.-F., {Zhang}, Q.-M., \& {Fang}, C. 2014, Research in Astronomy and Astrophysics, 14, 581, \dodoi{10.1088/1674-4527/14/5/007}

\bibitem[{{Zhou} {et~al.}(2023){Zhou}, {Li}, {Hong}, \& {Keppens}}]{zhou2023}
{Zhou}, Y., {Li}, X., {Hong}, J., \& {Keppens}, R. 2023, \aap, 675, A31, \dodoi{10.1051/0004-6361/202346004}

\bibitem[{Zirker {et~al.}(1994)Zirker, Engvold, \& Yi}]{zirker1994}
Zirker, J.~B., Engvold, O., \& Yi, Z. 1994, Solar physics, 150, 81

\end{thebibliography}
\bibliographystyle{aasjournal}



\end{document}